\documentclass[article,superscriptaddress]{revtex4}
\usepackage{amsmath}
\usepackage{amssymb}
\usepackage{graphicx}

\renewcommand{\a}{\alpha}

\newcommand{\e}{\epsilon}
\newcommand{\g}{\gamma}

\newcommand{\pe}{p}
\newcommand{\pes}{p_s}
\newcommand{\s}{\sigma}

\newcommand{\w}{\omega}
\newcommand{\tf}{\tilde{f}}
\newcommand{\E}{\mathbb{E}}
\renewcommand{\P}{\mathbb{P}}
\newcommand{\NN}{\mathbb{N}}

\newcommand{\N}{\mathcal{N}}
\newcommand{\C}{\mathcal{C}}
\newcommand{\A}{\mathcal{A}}
\newcommand{\X}{\mathcal{X}}
\renewcommand{\H}{\mathcal{H}}
\newcommand{\I}{\mathcal{I}}
\newcommand{\beq}{\begin{equation}}
\newcommand{\eeq}{\end{equation}}

\newcommand{\qi}{q_{\rm in}}
\newcommand{\qe}{q_{\rm env}}

\begin{document}

\title{The Value of Information for Populations in Varying Environments}

\author{Olivier Rivoire}
\affiliation{Laboratoire de Spectrom\'etrie Physique, CNRS \& Universit\'e Joseph Fourier, Grenoble, France}
\author{Stanislas Leibler}
\affiliation{Laboratory of Living Matter, The Rockefeller University, New York, New York, USA}
\affiliation{School of Natural Sciences, The Simons Center for Systems Biology, The Institute for Advanced Study, Princeton, New Jersey, USA\\ \ }

\date{\today}

\begin{abstract}
The notion of information pervades informal descriptions of biological systems, but formal treatments face the problem of defining a quantitative measure of information rooted in a concept of fitness, which is itself an elusive notion. Here, we present a model of population dynamics where this problem is amenable to a mathematical analysis. In the limit where any information about future environmental variations is common to the members of the population, our model is equivalent to known models of financial investment. In this case, the population can be interpreted as a portfolio of financial assets and previous analyses have shown that a key quantity of Shannon's communication theory, the mutual information, sets a fundamental limit on the value of information. We show that this bound can be violated when accounting for features that are irrelevant in finance but inherent to biological systems, such as the stochasticity present at the individual level. This leads us to generalize the measures of uncertainty and information usually encountered in information theory. 
\end{abstract}

\maketitle

\section{Introduction}

Information is a central concept in biology~\cite{MaynardSmith00,Jablonka02,Szostak03,Nurse08}, which many studies have sought to formalize~\cite{Quastler53,Rashevsky55,Atlan72,Berger03,Adami04,Taylor07,Polani09}. In this quest, Shannon's theory of communication~\cite{Shannon48} has always played an influential role. Originally, this theory was concerned with two basic problems: the problem of efficiently encoding signals, and the problem of reliably transmitting them through noisy channels. Shannon proposed a formal framework within which these questions could be addressed mathematically. By modeling information sources and communication channels in probabilistic terms, and by focusing on the asymptotic properties of long sequences of symbols, he established fundamental limits for the achievable rates of data compression and transmission~\cite{Shannon48,Shannon59}. By virtue of the abstract nature of the model, these limits hold irrespectively of the particular material implementation. Remarkably, the same quantity, the mutual information $I(X;Y)$, a function of two random variables $X$ and $Y$, emerges as a common measure of "information" in the solution of the two problems~\cite{CoverThomas91}. As for the related concept of entropy $H(X)=I(X;X)$, the definition of the mutual information can be axiomatized~\cite{Shannon48,Csiszar08}, which has lent support to the view that this quantity represents an universal and irrefutable measure of information. The emergence of the mutual information as a central quantity in problems of point-to-point communication however rests on specific assumptions, which have to be reexamined in any other instance where a concept of "information" is to be formalized~\cite{Shannon56}.\\

A class of problems where such a reexamination has led to identifying a different measure of information is constituted by the engineering problems of control. These problems share two essential features with biological systems:  information is processed for a "function", which confers value to the information, and feedback, whereby elements from the past are used to affect the present, is essential. Historically, these parallels between regulation in living organisms and control in engineered systems has underlaid the seminal works on control with feedback~\cite{RosenbluethWienerBigelow43}. It also motivated the influential development of cybernetics, which Wiener defined as "the science of control and communication, in the animal and the machine"~\cite{Wiener48}. A law formulated in the early days of cybernetics is thus the "law of requisite variety"~\cite{Ashby56,Ashby58}, which states that the value of information for control cannot exceed the limit set by the mutual information between a disturbance and its measurement (see also~\cite{TouchetteLLoyd00,TouchetteLloyd04}). The issue of quantifying information in systems of control has been revisited thoroughly since this law was proposed~\cite{Mitter01}. These analyses have concurred to establish the so-called directed information~\cite{Marko73,Massey90} as a measure of information more relevant than the mutual information when issues of feedback are involved. The directed information measures the causal dependence between two stochastic processes, in contrast with the mutual information which ignores any constraint of causality and only measures statistical correlations. Consistently with the law of requisite variety, the mutual information however appears as an upper bound for the value of information for control when the later is measured by a directed information.\\

The parallel between living organisms and engineered systems provides interesting insights but fails to account for two other essential features of living organisms: their organization into populations, and the need to evaluate performance in terms of "fitness", i.e., in terms of an appropriate measure of reproductive value. Viewing the problem of control from the standpoint of populations of reproducing individuals indeed introduces new options for coping with unpredictable variations of the environment. Most importantly, a "bet-hedging strategy"~\cite{SegerBrockmann87} can be implemented through the diversification of the population. An analogy with financial problems of risk management has been noticed many times~\cite{LewontinCohen69,Real80,Stearns00,Wagner03}, including from the perspective of information processing~\cite{Stephens89,BergstromLachmann04,KussellLeibler05,Donaldson10}. Both problems involve a growing population facing an unpredictable future: in the financial problem, the population is composed by the capital of an investor, which is distributed between different assets. These assets are analogous to the phenotypes of biological organisms, and may respond differently to different environmental perturbations. The problem of quantifying the value of information in this context was first analyzed by Kelly~\cite{Kelly56}, who found that the mutual information appeared as a natural measure. His results were later expanded~\cite{Breiman61,AlgoetCover88,BarronCover88,Cover98} showing that, in general, the relevant measure for the value of information must incorporate characteristic features of the individuals, such as their multiplication rates. A result analogous to the law of requisite variety however still holds: the value of the information that an investor may collect remains bounded by the mutual information between this information and the actual state of the environment (here the stock market)~\cite{BarronCover88}.\\

The analogy between biological populations and problems of financial investment has also its limitations. The main conceptual difference is that the financial problem is supervised by a goal-oriented investor, who centralizes the information and the decisions, while information processing is distributed between potentially independent individuals in biological populations. A first implication is that the biological problem may not correspond to an optimization at the population level, as it does by definition in finance. In any case, the justification of a criterion of optimality must involve a non-arbitrary objective function that emerges from the dynamics of the population instead of being a priori defined. The distributed nature of the biological problem also introduces a level of individual stochasticity that is absent in finance: even if every individual has the same sensor and has access to the same information, stochastic noise within each individual sensor can lead to the perception of non-identical signals. This aspect of the problem of information processing, which has not been previously examined from an information theoretic standpoint, also leads to a measure of the value of information that differs from the mutual information. In this case, the law of requisite variety may also be violated: the value of the relevant measure of information can exceed the value indicated by the mutual information. A population may thus effectively acquire, in a distributed form, a more accurate information than any of its members.\\

We shall discuss each of these points in the context of a mathematical model of growing populations in a varying environment. This model is defined in Sec.~\ref{sec:model} and its main elements are represented in Fig.~\ref{fig:scheme}. It deals with two types of biological information: the information inherited by an individual from its parents, and the information directly acquired from the environment. To exploit the analogies with the engineering problem of control and the financial problem of investment (see Table~\ref{fig:tablecorres}), we define and justify in Sec.~\ref{sec:fitness} a suitable "fitness function". Our presentation is then organized around three simplifying assumptions: assumption (A1) that individuals have no memory, assumption (A2) that individuals all perceive the same information from the environment, and assumption (A3) that only individuals perfectly adapted to their environment can survive. While under the conjunction of these three assumptions, the value of information is expressed by a mutual information (Sec.~\ref{sec:horseraces})~\cite{Kelly56}, relaxing any of these assumptions exposes a different limitation of this measure of information. Relaxing (A1) introduces the possibility of feedback, in which case constraints of causality not accounted for by the mutual information need to be incorporated (Sec.~\ref{sec:dirinfo})~\cite{PermuterKim08}. Relaxing (A2) introduces the possibility for individuals to perceive different signals from their common environment, which also requires generalizing the mutual information (Sec.~\ref{sec:indiv}). Finally, relaxing (A3) introduces the possibility of different environmental states having non-exclusive "meaning", where the source of meaning, encapsulated in the values of the multiplication rates of the individuals, needs to be taken explicitly into account in the measure of information (Sec.~\ref{sec:f})~\cite{CoverThomas91}. Different expressions for quantifying the value of information are thus obtained, which are summarized in Table~\ref{fig:tablesum}.\\

Besides the question of quantifying the value of information, our model also addresses a second question, the question of characterizing the evolutionary stable strategies that optimize fitness. We shall show that, under the assumptions (A2) and (A3), these strategies amount to a Bayesian computation, as conjectured for instance in~\cite{PerkinsSwain09}. When these assumptions are not satisfied, however, we find that population-level features can make the implementation of a Bayesian computation irrelevant.

\section{Model}\label{sec:model}

Our approach to investigating the nature and value of information in biological systems is based on an abstract mathematical model. Expressions for the value of information will result from analyzing this model, both at the individual level of organisms, at which the model is defined, and at the population level. Specifically, our model seeks to incorporate the following features, which appear to be commonly shared by all living organisms:\\

\begin{tabular}{ r l  }
(i) & Living organisms change (as a result of development, phenotypic plasticity, learning,\dots);\\
(ii) & Living organisms can generate other living organisms;\\
(iii) & The faculties (i) and (ii) are affected by the state of the organism and the state of its environment;\\
(iv) & The environment of living organisms varies.
\end{tabular}\\

The issue of regulation arises when constraints are present which prevent the organisms from perfectly anticipating environmental changes. Here, we focus on constraints due to limited information (see Sec.~\ref{sec:general} for extensions):\\

\ (v) Changes within a living organism take place in absence of complete information about the forthcoming environmental states that will affect survival and reproduction.\\

To account for (iv), the environment is described by a discrete-time and discrete-state Markov chain, with transition matrix $\pe(x_t|x_{t-1})$. This Markov chain is assumed to be stationary and ergodic. We shall expend on the notion of ergodicity in Sec.~\ref{sec:fitness}, but, in essence, it requires that any environmental state can be reached from any other state in finite time and with finite probability~\cite{KarlinTaylor75}. Ergodic Markov chains tend asymptotically to an unique stationary distribution $\pes(x_t)$, irrespectively of their initial state, where $\pes(x_t)$ satisfies $\pes(x_t)=\sum_{x_{t-1}}\pe(x_t|x_{t-1})\pes(x_{t-1})$. We assume here that the environmental process is stationary. A particular case of interest is when the successive environmental states are uncorrelated and described by independently and identically distributed (i.i.d.) random variables, each having a probability $\pe(x_t)$, corresponding to $\pe(x_t|x_{t-1})=\pe(x_t)=\pes(x_t)$.\\

\begin{figure}[t]
\begin{center}
\includegraphics[width=.6\linewidth]{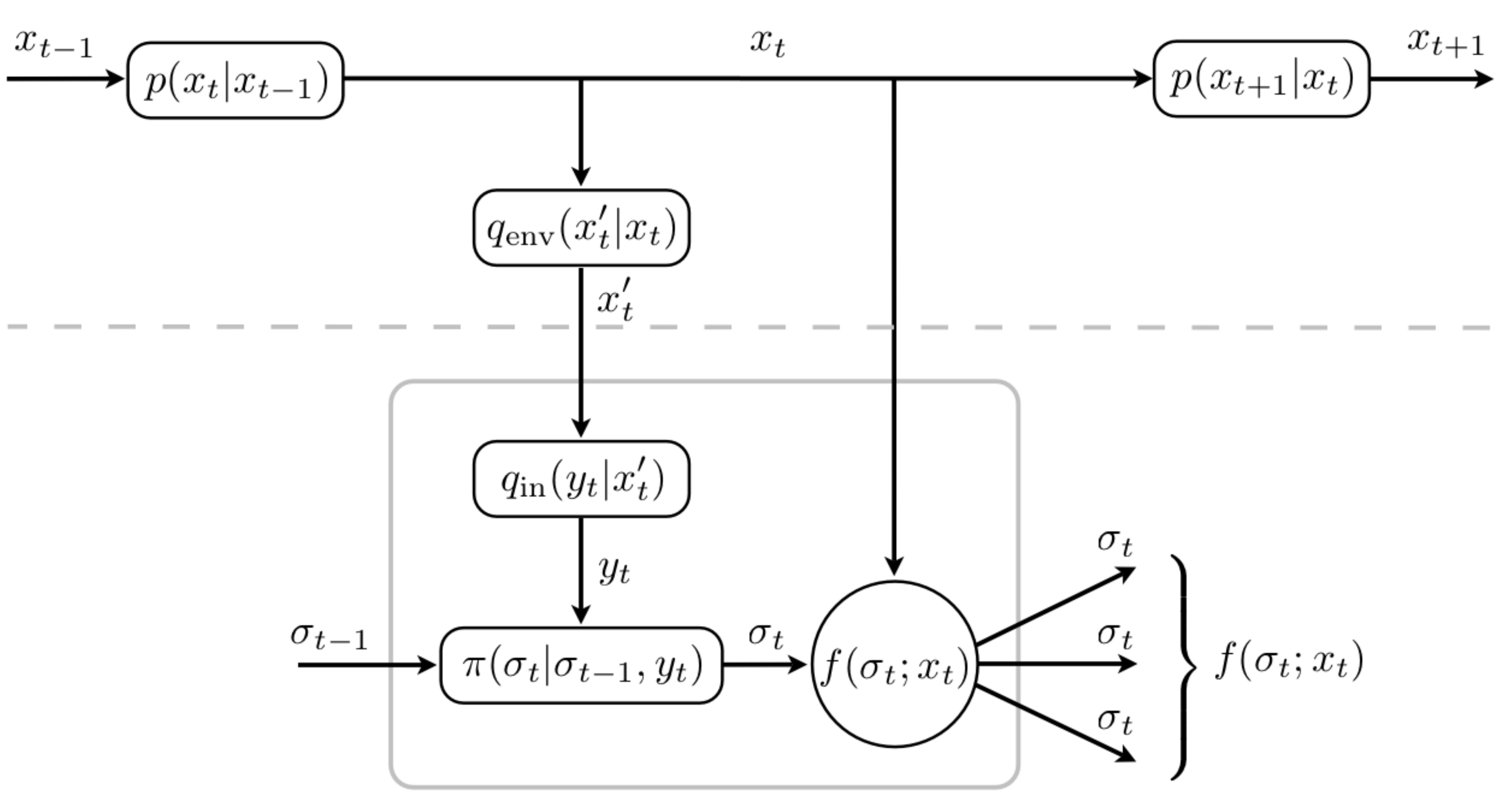}
\caption{Schematic representation of the relation between the environment (upper part) and an individual (lower part).The current environmental state $x_t$ affects the number $f(\s_t;x_t)$ of offsprings than an individual of type $\s_t$ generates (or, more generally, the expected number of offsprings given $\s_t$ and $x_t$). The type $\s_t$ is determined probabilistically by the transition matrix $\pi(\s_t|\s_{t-1},y_t)$ which depends on the ancestral type $\s_{t-1}$ and on the perception $y_t$ that the individual has of the environment. The signal $y_t$ derives from $x_t$ through two possibly noisy communication channels: an "environmental" channel $\qe(x'_t|x_t)$, which specifies a perceptible signal $x'_t$ common to the whole population, and an "individual" channel $\qi(y_t|x'_t)$, which specifies, independently for each individual, a perceived signal $y_t$. This second channel is noiseless in the financial interpretation of the model for which no stochasticity is present at the individual level; in this case $y_t=x'_t$, and the noiseless individual channel is denoted $\qi=\delta$ (see also Fig.~\ref{fig:trivialchannels}).}
\label{fig:scheme}
\end{center}
\end{figure}

Each individual organism is characterized by an internal state $\s_t$, to which we will refer as its current "type"; in general, it corresponds to a distinct phenotype, but may also be associated with a distinct genotype.  To account for (ii), the number $f(\s_t;x_t)$ of offsprings generated by an individual organism at time $t$ depends both on its type $\s_t$, and on the current state $x_t$ of the environment; in particular, the individual may die if $f(\s_t;x_t)=0$ or survive without reproducing if $f(\s_t;x_t)=1$. As a simplifying assumption, we assume here that all offsprings inherit the type $\s_t$ of their parent. More generally, a non-integer value of $f(\s_t;x_t)$ will represent the expected number of offsprings of an individual of type $\s_t$ in environment $x_t$; $f(\s_t;x_t)$ will therefore be called a multiplication rate. To account for (iii) and (v), the current type $\s_t$ can depend both on the ancestral type $\s_{t-1}$ of the individual, and on a signal $y_t$ derived from the environment $x_t$. Following the example of communication theory~\cite{Shannon48}, this dependence is described probabilistically, with a transition matrix $\pi(\s_t|\s_{t-1},y_t)$ giving the probability to end up in state $\s_t$ given $(\s_{t-1},y_t)$. In the language of information theory, such a transition matrix is also called a "communication channel", here with input $(\s_{t-1},y_t)$ and output $\s_t$; mathematically, it must satisfy two basic properties:
\beq
\pi(\s_t|\s_{t-1},y_t)\geq 0,\quad {\rm for\ all\ }\s_t,\s_{t-1},y_t,\quad{\rm and}\quad \sum_{\s_t}\pi(\s_t|\s_{t-1},y_t)=1\quad{\rm for\ all\ } \s_{t-1},y_t.
\eeq
The relation between the signal $y_t$ and its source $x_t$, is also specified probabilistically. To distinguish between the common and individual levels of stochasticity, we describe this relation with two consecutive communication channels (see Fig.~\ref{fig:scheme}): a first communication channel attached to the environment, $\qe(x'_t|x_t)$, whose output is a cue $x_t'$ common to all individuals in the population, followed by a second communication channel attached to each individual, $\qi(y_t|x'_t)$, whose output is the signal $y_t$. For instance, if considering a population of bacteria, $x_t$ may represent the chemicals constituting the medium at time $t$, $x'_t$ the subset of those chemicals for which the bacteria have a sensor, and $y_t$ the chemicals that a particular bacterium actually detects at time $t$, which may vary from bacteria to bacteria due to imperfect sensors. The difference between $x_t$, the environmental state affecting the multiplication rate $f(\s_t;x_t)$, and $x'_t$, the environmental cue, may also represent a delay between sensing and reproduction~\footnote{For instance, we may consider $x_t=(e_t,e_{t-1})$ and $x_t'=e_{t-1}$, with $e_t$ described by a Markov chain with transition matrix $b(e_t|e_{t-1})$ and $p(x_t|x_{t-1})=b(e_t|e_{t-1})$.}.\\

{\bf Equation for the conditional mean population size $\N_t$ --} The model is defined at the level of individual organisms, but selection may also act at the level of the population; for instance, a diversification between different types may confer an advantage when the environmental changes are unpredictable. An important implication is that the problem of regulation in a varying environment should not be treated by isolating an individual from the population. Here, the population is characterized by the numbers $Z_t(\s)$ of individuals of each type $\s$, which define a population vector $Z_t$ whose norm $|Z_t|\equiv\sum_\s Z_t(\s)$ is the total population size. This vector $Z_t$ is a random variable from two standpoints: it depends on the environmental sequence $\bar x\equiv((x_1,x_1'),\dots,(x_t,x'_t),\dots)$, and for a given $\bar x$, it is subject to the stochasticity at the individual level, generated through the transition matrices $\qi(y_t|x'_t)$ and $\pi(\s_t|\s_{t-1},y_t)$ (and possibly also through the fluctuations in the number of offsprings if $f(\s_t;x_t)$ represents a multiplication rate). We will use two different symbols for representing the two corresponding averages: $\langle Z_t(\s)\rangle$ for the average conditionally to the environmental sequence $\bar x$ 
, and $\E[\langle Z_t\rangle]$ for the average over environmental sequences as well.
Our analysis will focus on the conditional mean 
\beq
\N_t(\s)\equiv\langle Z_t(\s)\rangle\qquad\textrm{(average taken for a given $\bar x$)},
\eeq
which follows a simple recursion:
\beq\label{eq:dyn}
\N_t(\s_t)=f(\s_t;x_t)\sum_{\s_{t-1},y_t}\pi(\s_t|\s_{t-1};y_t)\ \qi(y_t|x'_t)\ \N_{t-1}(\s_{t-1}).
\eeq
This recursion can also be written with a vectorial notation:
\beq\label{eq:A}
\N_t=\A^{(t)} \N_{t-1},\quad\textrm{with}\quad \A_{\s'\s}^{(t)}\equiv f(\s';x_t)\sum_{y_t}\pi(\s'|\s;y_t)\ \qi(y_t|x'_t),
\eeq
where $\A^{(t)}$ is a shorthand for $\A^{(x_t,x'_t)}$. Here, the current environment $(x_t,x'_t)$ is a "quenched" variable, which is fixed independently of the dynamics of the population. From a mathematical standpoint, Eq.~\eqref{eq:A} indicates that studying $\N_t$ amounts to studying the product of random matrices $\A^{(t)}\A^{(t-1)}\dots \A^{(1)}$, which is function of the environmental sequence $\bar x$. In contrast to $Z_t$, $\N_t=\langle Z_t\rangle$ overlooks the discrete nature of the population, and thus fails to account for possible events of extinction; a population of discrete individuals is indeed not infinitely divisible, and the stochasticity of the process of reproduction may lead to $|Z_t|=0$ at some time $t$, after which any possibility of recovery is excluded. Remarkably however, the results presented in Sec.~\ref{sec:fitness} indicate that the basic asymptotic behavior of $|Z_t|$ can be derived from the properties of $|\N_t|$, which will justify that our analysis concentrates on Eq.~\eqref{eq:dyn}.\\

\begin{figure}[t]
\begin{minipage}[c]{.46\linewidth}
\begin{center} 
\renewcommand{\arraystretch}{1.5} 
\begin{tabular}{  c | c | c }
  & {\it acquired}  & {\it inherited} \\ 
\hline {\it population} & perceptible $x'_t$ & transition matrix $\pi$\\
\hline  \hspace{.2cm} {\it individual}  \hspace{.2cm} &  \hspace{.2cm} perceived $y_t$  \hspace{.2cm} &  \hspace{.2cm} type $\s_t$  \hspace{.5cm}
\end{tabular} 
\end{center}
\caption{\small Four different notions of "information" contained in the model. Information has two sources, the environment and the ancestor of the individual, corresponding to acquired or inherited information, and is defined at two levels, the individual or the population. The transition matrix $\pi(\s_t|\s_{t-1},y_t)$ may be viewed as information about the environment encoded in the organisms (see e.g. Eq.~\eqref{eq:propbetting}). If mutations could lead to the unreliable transmission of $\pi$, an extra level of description would be introduced, with subpopulations characterized by different values of $\pi$.\label{fig:info}}
\end{minipage} \hfill
\begin{minipage}[c]{.46\linewidth}
\centering
\includegraphics[width=\linewidth]{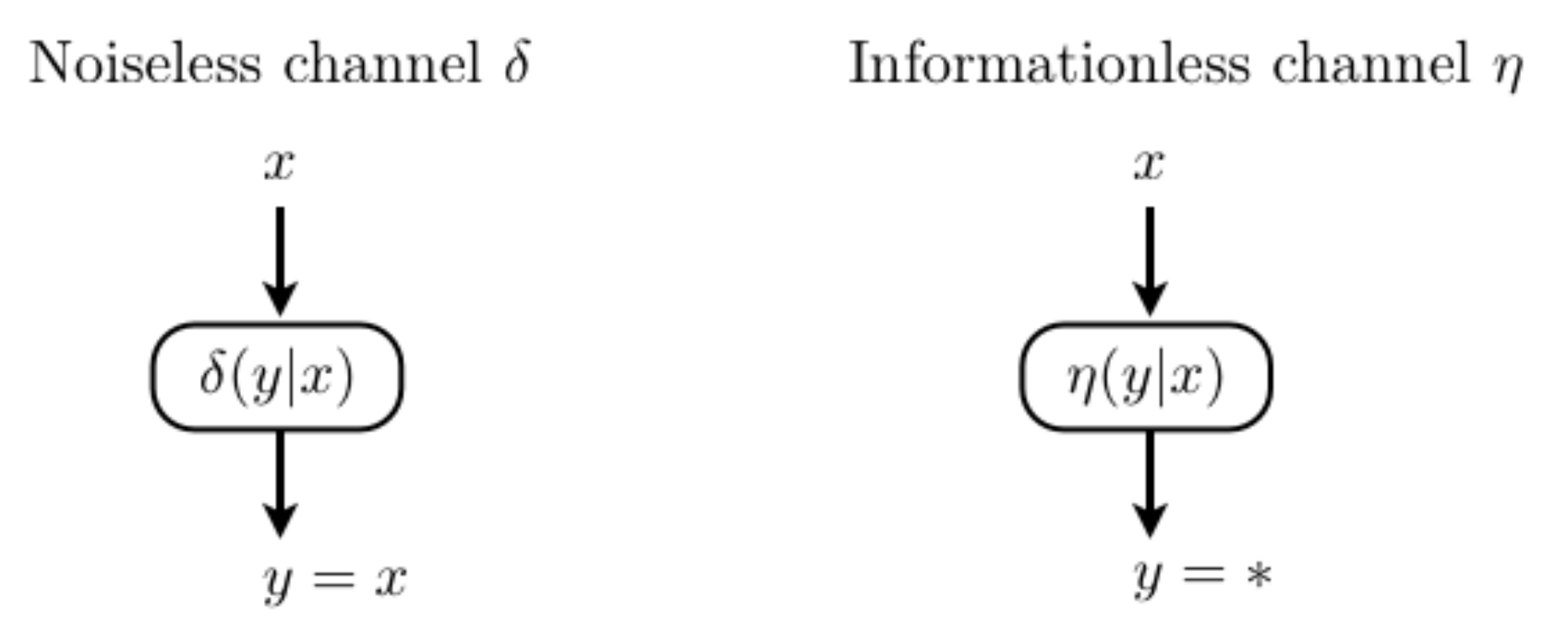}
\caption{Two trivial communication channels, the noiseless channel $\delta$ which transmits perfectly information, and the informationless channel $\eta$ which does not transmit any information. We have for instance $\qi=\delta$ in the financial interpretation of the model where no individual stochasticity is present, and $\qe=\eta$ in the case where decisions must be made in absence of any acquired information about the current environmental state (see Fig.~\ref{fig:channels} for less trivial examples of communication channels).}\label{fig:trivialchannels}
\end{minipage}
\end{figure}

{\bf Financial interpretation --} The decomposition of the channel of acquired information into an environmental channel $\qe(x'_t|x_t)$ and an individual channel $\qi(y_t|x'_t)$ is further motivated by the financial interpretation of our model, where only the environmental channel $\qe(x'_t|x_t)$ make sense~\footnote{An equivalent model could indeed have been defined without reference to $\qe$ by considering $\tilde q_{\rm in}(y_t|x_t)=\sum_{x'_t}\qi(y_t|x'_t)\qe(x'_t|x_t)$. The decomposition into $x'_t$ and $y_t$, although not unique, cannot be arbitrary, and it will provide us with interesting inequalities in Sec.~\ref{sec:indiv}.}. In this interpretation, $\N_t(\s_t)$ represents the number of currency units that an investor on the stock market invests in asset $\s_t$ on day $t$, $\pi(\s_t|\s_{t-1},y_t)$ represents the fraction of money transferred from asset $\s_{t-1}$ to asset $\s_t$, based possibly on some information $y_t$ available about the current state $x_t$ of the market, and $f(\s;x_t)$ represents the return of asset $\s$ on day $t$, a non-negative but non-necessarily integer quantity (see Table~\ref{fig:tablecorres}). Eqs.~\eqref{eq:dyn} and \eqref{eq:A} then describe the evolution of $\N_t$ in a scenario where the money is entirely reinvested every time. The essential difference with the biological case is the absence of stochasticity at the level of individuals, which are strictly equivalent currency units: $\pi(\s_t|\s_{t-1},y_t)$ results from the decision of an investor which centralizes the information used for manipulating each of the currency unit constituting the "population". This has two implications: (i) $y_t$ has to be common to the population, i.e., $\qi=\delta$ and as a result $y_t=x'_t$ (see Fig.~\ref{fig:trivialchannels}); (ii) the only source of stochasticity is the environment, which operates at the level of the population, i.e., $Z_t=\N_t$ (note however that the finite divisibility of the currency unit is not accounted for if considering only $\N_t$). In contrast, the necessity for biological populations to process information at the level of individual organisms introduces an extra level of stochasticity and heterogeneity, which underlies qualitative differences with problems of financial investment.\\

{\bf Two basic questions --} The transition matrix $\pi(\s_t|\s_{t-1};y_t)$ specifies the "strategy" for responding to the signals that individuals inherit and acquire. A basic problem is to provide a framework for estimating the relative performance of different strategies. In some particularly cases, the "best" strategy is clear: if perfect information is available, a sensible action is indeed for every individual to adopt at time $t$ the type $\s$ that maximizes $f(\s;x_t)$, thus leading to an homogeneous population. Perfect information correspond to noiseless channels, represented by the identity transition matrix $\delta$ such that $\delta(y_t|x_t)=1$ if $y_t=x_t$, and 0 otherwise (see Fig.~\ref{fig:trivialchannels}). In general, however, the communication channels $\qe$ and $\qi$ will reveal incomplete information about $x_t$, and a non-deterministic response, leading to a diversified population, may be more advantageous. Two basic questions thus arise:\\

\begin{table}
\renewcommand{\arraystretch}{1.8} 
\setlength{\tabcolsep}{25pt} 
 \begin{center}
 \begin{tabular}{c|c|c||c} 
 Biology & Finance &  Control theory  & Notation \\ 
\hline \hline
individual & currency unit & system  & -\\
\hline
population & capital & - & -\\
\hline
phenotype & asset & state & $\sigma_t$\\
\hline
environment & market & disturbance & $x_t$\\
\hline 
multiplication rate & return & 1 & $f(\s_t,x_t)$\\
\hline
acquired info. & side-information & feedforward info. & $y_t$\\
\hline
inherited info. & - & feedback info. &  $\sigma_{t-1}$\\
\hline
strategy & portfolio & policy & $\pi(\sigma_t|\sigma_{t-1},y_t)$\\
\hline
fitness & utility & loss-function & $\Lambda(\pi)$
 \end{tabular}
 \end{center}
 \caption{Correspondence between the terms used in biology, finance and control theory. The engineering problem is concerned with a single system, and therefore involves no notion of multiplication rate or population. The financial problem is defined for an agent supervising any information processing, and the notion of information inherited by the individuals has therefore no obvious counterpart.}\label{fig:tablecorres}
\end{table}

 \begin{tabular}{ r l  }
(Q1) & What strategy, i.e., choice of the transition matrix $\pi(\s_t|\s_{t-1},y_t)$, is the most advantageous?\\
(Q2) & What is the value of the information acquired through $\qe(x'_t|x_t)$ and $\qi(y_t|x'_t)$?
\end{tabular}\\ \\

Answering these two questions require defining a measure of "fitness", so as to give a precise meaning to the notions of "advantage" and "value". In decision theory, this usually involves the introduction of an {\it ad-hoc} loss-function~\footnote{From this standpoint, our model is related to the so-called partially observable Markov decision processes studied in the operations-research literature~\cite{Pack98}.}. We show however in the next section that a measure of adaptation emerges in the long-term limit, which defines a non-arbitrary fitness function.\\

{\bf Three simplifying assumptions --} As the model is not analytically solvable in its most general form, it is of interest to analyze it under several simplifying assumptions. Three simplifying assumptions will play a crucial role:\\

 \begin{tabular}{ r l  }
(A1) & No information is inherited between successive generations, i.e, $\pi(\s_t|\s_{t-1},y_t)=\pi(\s_t|y_t)$;\\
(A2) & Any information acquired from the environment is common to all members of the population, i.e., $\qi=\delta$;\\
(A3) & The multiplication rates have a diagonal form, i.e., $f(\s;x)=f(x)>0$ if $\s=x$, and $f(\s;x)=0$ otherwise.\\ 
\end{tabular}\\ \\

Inherited information becomes useful in presence of correlations between successive environmental states and assumption (A1) is therefore restrictive only when the environment is not i.i.d.. In the language of control theory, presented in Table~\ref{fig:tablecorres} and developed in Sec.~\ref{sec:general}, assumption (A1) corresponds to an open-loop mode of control where feedback is absent. Assumption (A2) amounts to restricting to models which can be interpreted in financial terms, with no fluctuation in the signals perceived by the individuals. Assumption (A3) describes the situation where in any environmental state, there is only one type able to survive; in particular, this assumption assumes that the number of environmental states is the same as the number of types for the individuals.
The model defined by the conjunction of the three assumptions plays a special role, because, as explained in Sec.~\ref{sec:horseraces}, the two questions (Q1) and (Q2) have simple answers in terms of the standard measures of uncertainty and information from communication theory, the entropy $H(X_t)$  and the mutual information $I(X_t;Y_t)$ where $X_t$ and $Y_t$ refers to the random variables associated with the environmental state $x_t$ and the signal $y_t$ ($H(X_t)$ and $I(X_t;Y_t)$ are defined below). As we shall show, relaxing any of these assumptions introduces generalizations of these two quantities. The models satisfying all three assumptions were also the first models of population growth to be analyzed from the standpoint of information theory~\cite{Kelly56}. These models were originally interpreted as models of gambling in horse races, with $f(x_t)$ viewed as the pay-off when horse $\s=x_t$ wins and $\pe(x_t|x_{t-1})$ as the probability for it to happen, given that horse $x_{t-1}$ won the previous race. Generalizations to models of investment in the stock market, involving relaxation of the assumptions (A1) and (A3) have subsequently been considered from the same standpoint~\cite{CoverThomas91}. The relevance of this approach to understanding the adaptive value of strategies of diversification and the value of information in biological populations has also been previously noticed~\cite{BergstromLachmann04,KussellLeibler05}, although always under the restrictive assumption (A2) that information is acquired with no individual stochasticity. 

\section{Fitness and optimization}\label{sec:fitness}

The question (Q1) of defining an optimal strategy for a game or a financial investment whose outcome is uncertain has a long history, dating back from the earliest days of probability theory. We review here some of the solutions that have been proposed in this context before turning to their relevance for biological populations. We start by assuming that the environmental process is i.i.d.~and that no information is acquired (formally $\qi=\eta$ where the informationless channel $\eta$ is defined in Fig.~\ref{fig:trivialchannels}). In a model with no correlations between successive environmental states, no gain can be expected from knowing the previous state, and we can assume without restriction that the optimal transition matrix $\pi(\sigma_t|\sigma_{t-1})$ is of the form $\pi(\sigma_t|\sigma_{t-1})=\pi(\sigma_t)$. Under these assumptions, we need only consider the total size of the population, $|\N_t|=\sum_\sigma\N_t(\sigma)$ rather than the population vector $\N_t$. Indeed, 
\beq
|\N_t|=A^{(t)}|\N_{t-1}|,\quad\textrm{with}\quad A^{(t)}\equiv A^{(t)}(\pi)\equiv\sum_{\sigma'} \A_{\sigma'\sigma}^{(t)}=\sum_{\sigma'} f(\s';x_t)\pi(\s'),
\eeq
With these notations, a population of initial size $|\N_0|$ acquires after $T$ time steps a size $|\N_T|$ determined by a product of $T$ scalar random variables:
\beq\label{N_T}
|\N_T|=A^{(T)}A^{(T-1)}\dots A^{(1)} |\N_0|.
\eeq 

{\bf Arithmetic mean --} The difficulty of defining an optimal strategy $\pi$ basically stems from the fact that $|\N_T|$ is a random variable whose value depends on the particular sequence of environments $(x_1,\dots,x_T)$: to each such sequence corresponds an optimal strategy $\pi$, but in general no strategy is optimal for every environmental sequence. A na\"ive solution would be to maximize the expected return. When the successive environmental states are independent, this corresponds to
\beq
\E [A^{(T)}A^{(T-1)}\dots A^{(1)}]= \left(\E[A^{(t)}]\right)^T,\quad\textrm{with}\quad \E[A^{(t)}]=\E[A^{(t)}(\pi)]=\sum_x \pe(x) \sum_\s f(\s;x_t)\pi(\s).
\eeq
$\E$ denotes here the expectation with respect to the fluctuations of the environment. This leads to selecting the portfolio maximizing the so-called arithmetic mean of the return,
\beq
\max_\pi\E[A^{(t)}(\pi)]\qquad \textrm{(max. arithmetic mean)}.
\eeq
This strategy may however be very risky, as illustrated by the following example: consider a horse race involving only two horses $a$ and $b$ having equal probability of winning, with returns given by $f(a;a)=3$, $f(b;a)=0$ when horse $a$ wins, and $f(a;b)=0$, $f(b;b)=1/2$ when horse $b$ wins. The expected return, $f(a;a)\pi(a)+f(b;b)\pi(b)$, where $\pi(a)+\pi(b)=1$, is clearly optimized by betting everything on horse $a$, i.e., $\pi(a)=1$ and $\pi(b)=0$. But following this strategy in a sequence of races where the gains are systematically reinvested almost surely leads to bankruptcy. Indeed, if horse $b$ ever wins, everything is lost, and this happens with probability $1-(1/2)^T$, which tends to 1 as $T$ increases. The maximum expected return is indeed optimal only when averaging over all possible sequences of outcomes, in which case the gain resulting from the only environmental sequence where $a$ never fail to win more than compensate for the loss experienced with all the other sequences of outcomes. When dealing with a single sequence of outcomes, such an average over different environmental sequences is however not relevant.\\

{\bf Expected utility --} An argument often given in the economic literature, which dates back from D. Bernoulli's analysis of the famous St Petersburg paradox~\cite{Bernoulli38}, is that the criterion based on the arithmetic mean fails to recognize that small losses may represent more ''utility'' for the gambler than large gains. According to this view, utility of losses and gains depends on the gambler, and may for instance vary with the initial wealth $|\N_0|$.  At any given time, each investor should be considered as having his own utility function $u$ that he seeks to optimize,
\beq
\max_\pi\E[u(A^{(t)}(\pi))]\qquad \textrm{(max. expected utility $u$)}.
\eeq
The choice of $u(x)$ is critical, since it quantifies the notion of risk. Based on the postulate that an increase in wealth should result in an increase in utility inversely proportionate to the quantity of goods already possessed, Bernoulli proposed $u(x)=\ln x$ as a sensible form of the utility function. In finance, where the problem arises when selecting a diversified portfolio of assets, the risk is often measured by the expected return variance. The return of a given asset $\s$ at time $t$ corresponds in our model to the multiplication rate $f(\s;x_t)$, and the expected return of a portfolio can be written vectorially as $\pi^\top R$ where $\pi$ is the vector of portfolio weights $\pi(\s)$, and $R$ is the vector of expected returns $R(\s)=\E[f(\s;x)]$. Following a proposition made by Markowitz~\cite{Markowitz52}, the risk is usually measured by $\pi^\top\Sigma\pi$ where $\Sigma(\s,\s')=\E[f(\s;x)f(\s';x)]-R(\s)R(\s')$ represents the covariance matrix of the returns. The portfolio vectors $\pi$ maximizing $\pi^\top R-\lambda\ \pi^\top\Sigma \pi$  then defines a family of efficient portfolios parametrized by $\lambda$, a parameter fixing the degree of risk that the investor is ready to undertake ($\lambda$ can also be interpreted as a Lagrange multiplier for the maximization of $\pi^\top R$ at fixed level of risk $\pi^\top\Sigma \pi$, or alternatively, for the minimization of the risk for a fixed expected return). Except for the fact that the covariance matrix $\Sigma$ is used rather than the correlation matrix $C(\s,\s')=\E[f(\s;x)f(\s';x)]$, Markowitz criterion is essentially similar to maximizing a quadratic utility function $u(x)=x-\lambda x^2$. This function may be viewed as the second-order approximation of a more general utility function, where the approximation is justified by the difficulty of estimating higher-oder moments of the returns. Despite their widespread use, criteria based on utility theory and its variants however present a fundamental problem: they are based on {\it ad-hoc} definitions of risk.\\

{\bf Geometric mean --} An independent line of inquiry, initiated by Kelly~\cite{Kelly56,Mosegaard05}, has promoted the optimization of the geometric mean as an objective criterion. It is based on the observation that if $|\N_T|$ is indeed a random variable whose value depends on the particular sequence of outcomes, for large $T$ most sequences lead to a common, typical, value of the compound return. This can be seen as resulting from the strong law of large numbers applied to $\ln (|\N_T|/|\N_0|)=\sum_t \ln A^{(t)}$, which, as a sum of i.i.d. random variables, satisfies
\beq\label{eq:proba1}
\lim_{t\to\infty}\frac{1}{t}\ln |\N_t|=\E[\ln A^{(t)}(\pi)]\quad \textrm{with probability 1}.
\eeq
This result motivates the maximum geometric mean return strategy,
\beq\label{eq:mgm}
\max_\pi\E[\ln A^{(t)}(\pi)]\qquad \textrm{(max. geometric mean)},
\eeq
This criterion is formally equivalent to optimizing a logarithmic utility function, $u(x)=\ln x$, as originally proposed by Bernoulli in the framework of utility theory~\cite{Bernoulli38}. From this standpoint, it may appear as an arbitrary criterion~\cite{Samuelson71}, but the argument given here does not rely on the notion of utility function: it relies instead on a fundamental mathematical result, the strong law of large numbers.\\

Strategies $\hat \pi$ corresponding to maximizing the geometric mean as in Eq.~\eqref{eq:mgm} have, besides Eq.~\eqref{eq:proba1}, a number of other attractive properties~\cite{Mosegaard05,CoverThomas91} (a hat over a quantity, such as $\hat \pi$, will always indicate an optimized value of the quantity). From a biological point of view, a particularly important property  of $\hat \pi$ is asymptotical optimality in an even stronger sense than indicated by Eq.~\eqref{eq:proba1}: $\hat \pi$ outperforms any other strategy $\pi$ (which may vary in time) for almost every sequence of outcomes~\cite{CoverThomas91}, i.e.,
\beq\label{eq:largenumbers}
\limsup_{t\to\infty}\frac{1}{t}\ln \frac{|\N_t(\pi)|}{|\N_t(\hat \pi)|}\leq 0\quad \textrm{with probability 1.}
\eeq
From a biological standpoint, the strategy $\hat \pi$ is an evolutionary stable strategy~\cite{MaynardSmithPrice73}: a population characterized by $\hat \pi$ cannot be outnumbered by a population with a different $\pi$. In other words, if one were to start with a variety of species characterized by different $\pi$, one would almost surely end up with a population dominated by the species with largest geometric mean $\E[\ln A^{(t)}(\pi)]$. This justifies using the growth rate, given by the geometric mean $\E[\ln A^{(t)}(\pi)]$, as an unambiguous measure of adaptation, or "fitness". Fitness is often informally defined as the expected number of descendants of an individual in a given environment~\cite{MillsBeatty79}, which, in our model, would correspond to $\E[f(\s;x)]$ if considering the descendants after one generation. In general, however, the definition of a fitness function must be supplemented with the references to an "horizon" $T$ and to a particular sequence of future environmental states~\cite{BeattyFinsen89}. In our model, the growth rate emerges as an unique measure of fitness when considering the long-term limit $T\to\infty$, but, if considering a finite "horizon", there may be a different strategy $\pi$ that outperforms $\hat \pi$; for instance, at the scale of a single time step, a better strategy may be to optimize the expected multiplication rate, which essentially amounts to an optimization of the arithmetic mean. Note also that our measure of fitness for long-term adaptation is not attached to a particular individual but rather to a trait propagated in a population, the trait defined by the strategy $\pi$. An implication of the fact that the fitness function is defined at the population level is that we should not seek to interpret the behavior of the members of the population in terms of the maximization of an individual utility function~\cite{Robson96}.\\

The conclusion that the growth rate, given by the geometric mean, is the relevant fitness function in the long-term extends to cases where the environmental process is stationary and ergodic, but not necessarily i.i.d.. For an arbitrary environmental processes, the growth rate of a population will indeed depend on the particular sequence of environmental states $x_1,x_2,\dots,x_t$ that arises. Stationary ergodic processes, however, benefit from a self-averaging property: particular realizations of such processes tend with time to share common statistical features - features that reproduce those obtained by averaging over many particular sequences; this property is also known as the asymptotic equipartition property in information theory, where it plays an equally fundamental role and underlies the choice of considering infinitely long sequences of symbols~\cite{CoverThomas91}. For independent environments, ergodicity amounts to the law of large numbers, which was the crucial argument leading to Eq.~\eqref{eq:proba1}: almost all long sequences comprise a same fraction $\pe(x)$ of each state $x$. More generally, assuming that the environmental process is stationary and ergodic, and that $\E[\max(0,\ln \A^{(t)}_{\s'\s})]<\infty$ for all $\s,\s'$, where $\A_{\s\s'}^{(t)}$ is defined as in Eq.~\eqref{eq:A}, it can be shown~\cite{FurstenbergKesten60,Kingman73} that the limit
\beq\label{eq:Lyap}
\Lambda_{\pe;f}^{(\qe,\qi)}(\pi)\equiv\lim_{t\to\infty}\frac{1}{t}\E\ln |\N_t|
\eeq 
exists, and that
\beq
\lim_{t\to\infty}\frac{1}{t}\ln |\N_t|=\Lambda_{\pe;f}^{(\qe,\qi)}(\pi)\quad \textrm{with probability 1.}
\eeq
No simple analytical formula is available for $\Lambda_{\pe;f}^{(\qe,\qi)}(\pi)$, also known as a Lyapunov exponent, in the most general case, but an important exception is in absence of inherited information, when $\pi(\s_t|\s_{t-1},y_t)=\pi(\s_t|y_t)$ [assumption (A1)], in which case
\beq\label{eq:closedformed}
\Lambda_{\pe;f}^{(\qe,\qi)}(\pi)=\sum_{x_t,x'_t}\qe(x_t'|x_t)\ \pes(x_t)\ \ln\left(\sum_{\s_t,y_t}f(\s_t;x_t)\ \pi(\s_t|y_t)\ \qi(y_t|x'_t)\right).
\eeq\\

{\bf Typical {\it vs} mean population sizes --} Importantly, not only $\Lambda_{\pe;f}^{(\qe,\qi)}(\pi)$ describes the growth rate of the conditional mean $\N_t=\langle Z_t\rangle$, but also, under fairly general conditions, the growth rate of the size $|Z_t|$ of a typical population. An essential condition, however, is that the population does not become extinct. The probability of survival of a population with an arbitrary initial composition can always be expressed in terms of the probabilities $Q(\s|\bar x)$ of extinction of a population starting from a single individual of type $\s$: 
if starting with $\N_0(\s)$ individuals in each state $\s$, the probability of survival is indeed $\prod_\s(1-Q(\s|\bar x))^{\N_0(\s)}$, because each individual generates its own independent subpopulation. Here, we assume that either all the types have a non-zero probability to survive, i.e., $\P(Q(\s|\bar x)<1,\forall\s)=1$, or none of them survive, i.e., $\P(Q(\s|\bar x)=1,\forall\s)=1$; if this is not the case, we can always ignore the types that inevitably become extinct. Under this condition of regularity and a further technical condition of stability presented in appendix~\ref{SI:mbpre}, the following classification theorem holds~\cite{Tanny81}:
\beq\label{eq:tanny}
\begin{split}
{\rm (i) \ } & \Lambda_{\pe;f}^{(\qe,\qi)}(\pi)< 0 \implies \P[Q(\s|\bar x)=1,\forall \s]=1;\\
{\rm (ii)\ } & \Lambda_{\pe;f}^{(\qe,\qi)}(\pi)>0\implies \lim_{t\to\infty}\frac{1}{t}\ln |Z_t|=\Lambda_{\pe;f}^{(\qe,\qi)}(\pi)\quad\textrm{almost surely conditionally to non-extinction}.
\end{split}
\eeq
The second case where there is a non-zero probability of non-extinction is known as the supercritical case and is obviously the one of interest here. Remarkably, this theorem indicates that the growth of branching processes is controlled by the properties of the product of random matrices $\A^{(t)}\dots \A^{(1)}$ which governs the evolution of the conditional mean $\N_t$. Even the condition of stability, detailed in appendix~\ref{SI:mbpre}, bears on properties of this product: it basically requires that its columns all grow at a same rate so as to prevent too large fluctuations in the population size. Also note that both this condition of stability and the other condition of regularity relative to the probability of extinction are trivially satisfied for a single-type population, to which our model can be reduced in absence of inherited information, when $\pi(\s_t|\s_{t-1},y_t)=\pi(\s_t|y_t)$ [assumption (A1)], by noticing that a recursion can be written directly for $|N_t|=\sum_\s N_t(\s)$, as for instance in Eq.~\eqref{N_T}.\\

{\bf Reformulation of the two basic questions --} Based on the mathematical results presented in this section, the questions (Q1) and (Q2) introduced previously can be stated formally.  Taking the long-term growth rate $\Lambda_{\pe;f}^{(\qe,\qi)}(\pi)$ as a measure of fitness, (Q1) becomes the problem of finding a matrix $\hat \pi$ that maximizes it for given parameters $\pe$, $f$, $\qe$ and $\qi$ (while the optimal growth rate  $\hat\Lambda_{\pe;f}^{(\qe,\qi)}$ is unique, "the" optimal strategy $\hat \pi$ may not be). Based on the same principle, (Q2) becomes the problem of estimating $\hat \Lambda_{\pe;f}^{(\qe,\qi)}-\hat \Lambda_{\pe;f}^{(\eta,\eta)}$, where $\hat \Lambda_{\pe;f}^{(\qe,\qi)}=\Lambda_{\pe;f}^{(\qe,\qi)}(\hat \pi)$ denotes the optimal growth rate in presence of the channels $\qe$ and $\qi$, and $\hat \Lambda_{\pe;f}^{(\eta,\eta)}$ the optimal growth rate in their absence ($\eta$ denotes an informationless channel as in Fig.~\ref{fig:trivialchannels}). (Q1) and (Q2) thus amount to estimating the two following quantities:\\

 \begin{tabular}{r l l}
(Q1) & $\quad\hat \pi\equiv\arg\max_\pi\Lambda_{\pe;f}^{(\qe,\qi)}(\pi)\quad$ & (optimal strategy);\\
(Q2) & $\quad\hat \Lambda_{\pe;f}^{(\qe,\qi)}-\hat \Lambda_{\pe;f}^{(\eta,\eta)}\quad$ & (value of the information conveyed by $\qe$ and $\qi$).
\end{tabular}\\ \\

In the next section, we show that under the assumptions (A1), (A2) and (A3), the cost of uncertainty, defined as $\hat \Lambda_{\pe;f}^{(\delta,\delta)}-\hat \Lambda_{\pe;f}^{(\qe,\qi)}$ where  $\delta$ denotes a noiseless channel as in Fig.~\ref{fig:trivialchannels}, and the value of acquired information, defined as $\hat \Lambda_{\pe;f}^{(\qe,\qi)}-\hat \Lambda_{\pe;f}^{(\eta,\eta)}$ where $\eta$ denotes an informationless channel as in Fig.~\ref{fig:trivialchannels}, correspond respectively to the conditional entropy $H(X_t|Y_t)$ and the mutual information $I(X_t;Y_t)$. In Sec.~\ref{sec:dirinfo} and \ref{sec:indiv}, we show that upon relaxing the assumptions (A1) or (A2), the cost of uncertainty and value of acquired information are still independent of the multiplication matrix $f(\s;x)$, and thus define two quantities $H_\pe^{(\qe,\qi)}$ and $I_\pe^{(\qe,\qi)}$ that generalize the notions of conditional entropy and mutual information. Finally, we show in Sec.~\ref{sec:f}, that, in absence of any assumption, the statistical quantities $I_\pe^{(\qe,\qi)}$ and $H_\pe^{(\qe,\qi)}$ are bounds for the cost of uncertainty $\hat \Lambda_{\pe;f}^{(\delta,\delta)}-\hat \Lambda_{\pe;f}^{(\qe,\qi)}$ and the value of acquired information $\hat \Lambda_{\pe;f}^{(\qe,\qi)}-\hat \Lambda_{\pe;f}^{(\eta,\eta)}$ respectively. These results are summarized in Table~\ref{fig:tablesum}.

\section{Kelly's horse races}\label{sec:horseraces}

As originally shown by Kelly~\cite{Kelly56}, under the joint assumptions (A1), (A2) and (A3) stated in Sec.~\ref{sec:model}, a simple connection is found between the long-term growth rate and information theoretic quantities. We start by assuming that, in addition to the restrictions imposed by (A1), (A2) and (A3), the environment is i.i.d.~with probability $p(x_t)$ for the environmental states.\\

{\bf Value of information and cost of uncertainty in absence of acquired information --} In absence of acquired information ($\qe=\eta$), the long-term growth rate is given by Eq.~\eqref{eq:closedformed}:
\beq
\Lambda_{\pe;f}^{(\eta,\delta)}(\pi)=\sum_x \pe(x)\ln \left(f(x)\pi(x)\right).
\eeq
Taking into account the constraint $\sum_x\pi(x)=1$ with Lagrange multipliers, the answer to (Q1) is found to be the optimal strategy $\hat \pi$ given by
\beq\label{eq:propbetting}
\hat \pi(x)=\pe(x),\quad\forall x.
\eeq
This strategy is called proportional betting and has the remarkable property of not depending on the values of the returns $f(x)$. It yields an optimal growth rate that can be broken down in two terms:
\beq\label{eq:lambda}
\hat\Lambda_{\pe;f}^{(\eta,\delta)}=\sum_x\pe(x)\ln f(x)+\sum_x\pe(x)\ln \pe(x)=\hat \Lambda_{\pe;f}^{(\delta,\delta)}-H_\pe^{(\eta,\delta)}.
\eeq
The first term, $\hat \Lambda_{\pe;f}^{(\delta,\delta)}=\E[\ln f]=\sum_x\pe(x)\ln f(x)$, corresponds to the best conceivable growth rate in a typical sequence of races: it is achieved if the gambler knows in advance which horse is going to win and bets all his money on it. The second term, which is independent of $f$,
\beq
H_\pe^{(\eta,\delta)}\equiv \hat \Lambda_{\pe;f}^{(\delta,\delta)}-\hat\Lambda_{\pe;f}^{(\eta,\delta)}=-\sum_x\pe(x)\ln \pe(x)
\eeq
corresponds to Shannon's entropy for the random variable $X_t$, usually denoted $H(X_t)$ or $H(\pe)$ (we follow the common usage of representing by $X_t$ the random variable and by $x_t$ one of its values). The entropy quantifies the cost of uncertainty when the frequencies $\pe(x)$ are known but not the particular sequence of outcomes that occurs. Since $\hat \pi=\pe$, a good gambler must have a good estimate of the environmental distribution $\pe$. From the biological standpoint, a population well-adapted to a varying environment must, in this model, have evolved an "internal model of the environment" that encodes its statistical properties~\cite{KussellLeibler05}; in this sense, the matrix $\pi$ can be viewed as information about the environment that is common knowledge in the population (see Fig.~\ref{fig:info}).\\

{\bf Origins of the entropy --} The entropy $H_\pe^{(\eta,\delta)}$ appears in source coding theory as the optimal rate of lossless compression for the memoryless source $\pe$~\cite{Shannon48}. To understand why the same quantity occurs in the two problems, consider a sequence of $T$ environmental states: if there are $n$ possible states, the number of such sequences in $n^T=e^{T\ln n}$, and $\ln n$, the rate at which the number of possible sequences increases with $T$, provides a first plausible measure of uncertainty. This measure, originally proposed by Hartley~\cite{Hartley28}, does not account for the fact that some states may be less probable than others, thus effectively reducing the uncertainty. If $\pe(1),\dots,\pe(n)$ represent the probabilities of the $n$ different states, the law of large numbers indeed indicates that, almost surely, long environmental sequences are in state $x$ a fraction $\pe(x)$ of the time. The entropy $H_\pe^{(\eta,\delta)}=-\sum_x\pe(x)\ln\pe(x)$ corresponds to the rate of increase of the number of these typical sequences. The number of typical sequences is indeed $\Upsilon_\pe^{(T)}\equiv T!/[(\pe(1)T)!\dots(\pe(n)T)!]$ which, for $T\to\infty$, satisfies $(\ln \Upsilon_\pe^{(T)})/T\to-\sum_x\pe(x)\ln\pe(x)=H_\pe^{(\eta,\delta)}$. Since the typical sequences are all equiprobable, the entropy also characterizes the probability $e^{-TH_\pe^{(\eta,\delta)}}$ of observing a particular typical sequence; this property of asymptotic equipartition, which generalizes beyond i.i.d.~processes, is central to information theory~\cite{CoverThomas91}. The entropy satisfies $0\leq H_\pe^{(\eta,\delta)}\leq \ln n$, with $H_\pe^{(\eta,\delta)}=\ln n$ if and only if no reduction of uncertainty can be gained from the fact that some states are less probable than others, which is the case only when all states are equiprobable, i.e., $\pe(x)=1/n$ for all $x=1,\dots,n$. In the other extreme case where only one state can occur, say $\pe(1)=1$, the entropy takes its minimal value $H_\pe^{(\eta,\delta)}=0$, corresponding to an absence of uncertainty~\footnote{$\ln\pe(x)$ is not defined when $\pe(x)=0$, but $\pe(x)\ln\pe(x)\to 0$ when $\pe(x)\to 0$, and it is therefore understood in the definition of $H_\pe^{(\eta,\delta)}$ that $\pe(x)\ln\pe(x)=0$ whenever $\pe(x)=0$.}.\\

{\bf Cost of non-optimal strategies --} If the frequencies $\pe(x)$ are not estimated correctly by the gambler, suggesting a suboptimal strategy $\pi\neq \hat \pi$, an additional cost is incurred,
\beq\label{eq:D}
\Lambda_{\pe;f}^{(\eta,\delta)}(\pi)=\hat\Lambda_{\pe;f}^{(\eta,\delta)}(\hat \pi)-D(\hat \pi\|\pi).
\eeq
This cost involves another quantity playing a fundamental role in communication theory~\cite{CoverThomas91}, the so-called relative entropy, or Kullback-Leibler divergence, which is defined by 
\beq
D(\hat\pi\| \pi)\equiv\sum_x\hat\pi(x)\ln \frac{\hat\pi(x)}{\pi(x)}.
\eeq
It measures the deviation of the distribution $\hat \pi$ from the distribution $\pi$ and obeys the inequality $D(\pi\|\hat \pi)\geq 0$, with equality if and only if $\pi(x)=\hat \pi(x)$ for all $x$.\\ 

{\bf Value of information and cost of uncertainty in presence of acquired information --} We now assume that an information $y_t$ is available about the outcome $x_t$ of the race, through an external communication channel characterized by the transition matrix $\qe(y_t|x_t)$ (here $\qi=\delta$ and hence $x'_t=y_t$). The strategy $\pi$ can now depend on the signal $y_t$ with $\pi(\s_t|y_t)$ denoting the fraction of wealth bet on $\s_t$. For instance, there may be $n$ possible signals, in which case, no side-information would correspond to $\qe(y_t|x_t)=1/n$, and perfect side-information to $\qe(y_t|x_t)=1$ if $y_t=x_t$ and 0 otherwise. In general, some noise may cause $\qe(y_t|x_t)$ to be non-zero even if $y_t\neq x_t$. The expression for the growth rate is now
\beq
\Lambda_{\pe;f}^{(\qe,\delta)}(\pi)=\sum_{x,y}\qe(y|x)\pe(x)\ln \left(f(x) \pi(x|y)\right),
\eeq
where $\qe(y|x)\pe(x)$ represents the joint probability $\P_{X_t,Y_t}(x,y)$ that the environmental state is $x$ and the perceived signal is $y$. By conditioning with respect to the received signal $y$, the problem can be reduced to the case with no information: 
\beq\label{eq:argcond}
\Lambda_{\pe;f}^{(\qe,\delta)}(\pi)=\sum_{x,y}\P_{X_t,Y_t}(x,y)\ln (f(x) \pi(x|y))=\sum_y\P_{Y_t}(y)\left[\sum_x\P_{X_t|Y_t}(x|y)\ln (f(x) \pi(x|y))\right].
\eeq
For any given $y$, the optimization problem is therefore solved as before, with $\P_{X_t}(x)=\pe(x)$ replaced by $\P_{X_t|Y_t}(x|y)$. The optimal strategy, i.e., the answer to (Q1), is thus "conditional proportional betting":
\beq\label{eq:stratB}
\hat\pi(\s|y)=\P_{X_t|Y_t}(\s|y)=\frac{\P_{Y_t|X_t}(y|\s)\ \P_{X_t}(\s)}{\P_{Y_t}(y)}=\frac{\qe(y|\s)\ \pe(\s)}{\sum_{\s'} \qe(y|\s')\ \pe(\s')}.
\eeq
It exactly amounts to a Bayesian computation~\cite{PerkinsSwain09}.
The optimal value of the growth rate can again be broken down in two terms
\beq
\hat\Lambda_{\pe;f}^{(\qe,\delta)}=\hat\Lambda_{\pe;f}^{(\delta,\delta)}-H_\pe^{(\qe,\delta)}.
\eeq
The second term, $H_\pe^{(\qe,\delta)}$, is a generalization of the entropy $H_\pe^{(\eta,\delta)}$ known as the conditional entropy, usually denoted $H(X_t|Y_t)$ in communication theory~\cite{CoverThomas91}. It measures the residual unpredictability of $X_t$ given $Y_t$ and is given by
\beq\label{eq:HXY}
H(X_t|Y_t)=\sum_y\P_{Y_t}(y)\ H(X_t|Y_t=y)=-\sum_{x,y}\P_{X_t,Y_t}(x,y)\ \ln \P_{X_t|Y_t}(x|y). 
\eeq
With perfect side-information, $Y_t=X_t$, and the entropic cost is eliminated, $H(X_t|X_t)=0$, leaving only $\hat\Lambda_{\pe;f}^{(\delta,\delta)}=\E[\ln f]$. The gain in predictability due to the signal, i.e., the answer to question (Q2), is obtained by comparing the situations with and without side-information,
\beq
I_\pe^{(\qe,\delta)}\equiv \hat\Lambda_\pe^{(\qe,\delta)}-\hat\Lambda_\pe^{(\eta,\delta)}=H_\pe^{(\eta,\delta)}-H_\pe^{(\qe,\delta)}=H(X_t)-H(X_t|Y_t)\equiv I(X_t;Y_t).
\eeq
The quantity $I_\pe^{(\qe,\delta)}=I(X_t;Y_t)$ is another important measure of information in communication theory, the mutual information~\cite{CoverThomas91}. It appears in channel coding theory, the theory of reliable transmission through noisy channels~\cite{Shannon48}, where the capacity of the noisy channel $\qe$ is given by $C^{(\qe)}=\max_\pe I_\pe^{(\qe,\delta)}$, and in rate-distortion theory, the theory of lossy data compression~\cite{Shannon59}, where the optimal compression rate to describe a source $\pe$ within a mean distortion $D$ is given by $R_\pe(D)=\min_{\qe}\{ I_\pe^{(\qe,\delta)}:\E[d(x,y)]\leq D\}$, where $\E[d(x,y)]=\sum_{x,y}\qe(y|x)\pe(x)d(x,y)$ is the mean distortion for a given distance function $d(x,y)$ between the symbol $x$ from the original data and the symbol $y$ from the compressed data. When $X=Y$, the mutual information $I(X;X)$ is nothing but the entropy $H(X)$. The mutual information between two random variables $X$ and $Y$ can also be expressed as $I(X;Y)=H(X)+H(Y)-H(X;Y)$, or as the relative entropy between the joint distribution of $(X,Y)$ and the product of their marginal distributions, i.e., $I(X;Y)=D(\P_{X,Y}\|\P_X\P_Y)$. It shows that $I(X;Y)$ is a symmetric function of its variables, and is always non-negative, with $I(X;Y)=0$ if and only if $X$ and $Y$ are independent, i.e., $\P_{X,Y}(x,y)=\P_X(x)\P_Y(y)$. The mutual information is thus a measure of statistical dependence between random variables.\\

{\bf Conclusion --} We assumed so far that the environment was i.i.d.~but under the assumption (A1) that no information can be inherited, the results of this section can be simply extended to Markov environments, and more generally to ergodic and stationary environmental processes, by simply replacing $\pe(x_t)$ by the stationary distribution $\pes (x_t)$ of the environmental process. To sum up, the growth rate for a model of horse races, defined by the assumptions (A1), (A2), (A3),  can be decomposed as
\beq
\Lambda_{\pe;f}^{(\qe,\delta)}(\pi)=\hat\Lambda_{\pe;f}^{(\delta,\delta)}-H_\pe^{(\qe,\delta)}-D(\hat \pi|| \pi),
\eeq
or, equivalently, as
\beq\label{3entrop}
\Lambda_{\pe;f}^{(\qe,\delta)}(\pi)=\hat\Lambda_{\pe;f}^{(\delta,\delta)}-H_\pe^{(\eta,\delta)}+I_\pe^{(\qe,\delta)}-D(\hat \pi||\pi).
\eeq
$\hat\Lambda_{\pe;f}^{(\delta,\delta)}=\E[\ln f]$ represents the optimal growth rate with perfect information, and $D(\hat\pi||\pi)$ the cost for following a strategy $\pi$ differing from the optimal strategy $\hat \pi$, with $D(\hat\pi||\pi)=0$ if and only if $\pi=\hat\pi$. The first expression makes apparent the cost of uncertainty $H_\pe^{(\qe,\delta)}$, which corresponds here to a conditional entropy:
\beq\label{eq:conde}
H_\pe^{(\qe,\delta)}=H(X_t|Y_t),
\eeq
where $\P_{X_t,Y_t}(x_t,y_t)=\qe(y_t|x_t)p_s(x_t)$.
The second expression introduces $H_\pe^{(\eta,\delta)}=H(X_t)$, the entropy of the environmental variable $X_t$, for which $\P_{X_t}(x_t)=\pes(x_t)$, and it makes explicit the value of acquired information $I_\pe^{(\qe,\delta)}$, which corresponds to the mutual information between the environment $X_t$ and the acquired information $Y_t=X'_t$:
\beq
I_\pe^{(\qe,\delta)}=I(X_t;Y_t).
\eeq
In the next three sections, we examine how these relations are modified when relaxing any of the assumptions (A1), (A2) and (A3) on which they rely. 

\section{Causal constraints and inherited information}\label{sec:dirinfo}

We first consider the consequences of relaxing the assumption (A1) by allowing information to be inherited. Under the assumptions (A2) and (A3) that the model still admits an interpretation in terms of horse races, so that in particular $\s_{t-1}=x_{t-1}$, the argument used to derive Eq.~\eqref{eq:argcond} can be invoked to infer that the Bayesian strategy, given by $\hat\pi(\s|x_{t-1},y_t)=\P_{X_t|X_{t-1},Y_t}(\s|x_{t-1},y_t)$, is optimal~\cite{CoverThomas91}, with an associated cost of uncertainty independent of $f$ and given by
\beq\label{eq:noH1}
H_\pe^{(\qe,\delta)}\equiv\hat\Lambda_{\pe;f}^{(\delta,\delta)}-\hat\Lambda_{\pe;f}^{(\qe,\delta)}=H(X_t|X_{t-1},Y_t).
\eeq
Here, following the definition of Eq.~\eqref{eq:HXY}, $H(X_t|X_{t-1},Y_t)$ is given by
\beq
H(X_t|X_{t-1},Y_t)=-\sum_{x_t,x_{t-1},y_t}\qe(y_t|x_t)\pe(x_t|x_{t-1})\pes(x_{t-1})\ln \left(\frac{\qe(y_t|x_t)\pe(x_t|x_{t-1})}{\sum_z\qe(y_t|z)\pe(z|x_{t-1})}\right).
\eeq

{\bf Value of information and cost of uncertainty in absence of acquired information --} In absence of acquired information ($\qe=\eta$), the uncertainty cost reduces to $H_\pe^{(\eta,\delta)}=H(X_t|X_{t-1})$. This cost is smaller than the uncertainty cost incurred in absence of inherited information, which was shown in the previous section to be $H(X_t)$. The difference is the mutual information $I(X_t;X_{t-1})=H(X_t)-H(X_t|X_{t-1})$, which thus quantifies the value of inherited information in this context.\\

The uncertainty cost $H_\pe^{(\eta,\delta)}=H(X_t|X_{t-1})$ can also be interpreted as the entropy rate $\H(X)$ of the environmental process: denoting $X^T\equiv (X_1,\dots,X_T)$, the entropy rate of the environmental process $X$ is generally defined by $\H(X)=\lim_{T\to\infty}H(X^T)/T$~\cite{CoverThomas91}. The limit always exists for a stationary ergodic process and it corresponds to $\H(X)=H(X_t|X_{t-1})$ for a Markov chain, and $\H(X)=H(X_t)$ for an i.i.d.~process.\\

{\bf Value of information in presence of acquired information --} In presence of both acquired and inherited information, it follows from Eq.~\eqref{eq:noH1} that the value of acquired information is given by
\beq\label{eq:condMI}
I_\pe^{(\qe,\delta)}\equiv H_{\pe}^{(\eta,\delta)}-H_{\pe}^{(\qe,\delta)}=H(X_t|X_{t-1})-H(X_t|X_{t-1},Y_t)\equiv I(X_t;Y_t|X_{t-1}),
\eeq
where the last equality defines the conditional mutual information $I(X_t;Y_t|X_{t-1})$. Using a conditioned version of the general relation $I(X;Y)=H(X)-H(X|Y)=H(Y)-H(Y|X)$, this conditional mutual information can also be written $I(X_t;Y_t|X_{t-1})=H(Y_t|X_{t-1})-H(Y_t|X_t)$. It is instructive to compare this quantity with the rate of mutual information between the processes $X$ and $Y$, which is defined by $\I(X;Y)=\lim_{T\to\infty}I(X^T;Y^T)/T$, where we use again the notations $Y^T=(Y_1,\dots,Y_T)$ and $X^T=(X_1,\dots,X_T)$. Given that $I(X^T;Y^T)=H(Y^T)-H(Y^T|X^T)$, the rate of mutual information corresponds here to $\I(X;Y)=\H(Y)-H(Y_t|X_t)$, where $\H(Y)$ represents the entropy rate for the process $Y$ (as an hidden Markov chain derived from a stationary ergodic chain, $Y$ has indeed a well-defined entropy rate). From $H(Y_t|X_t)\leq\H(Y)$, it follows that $I_\pe^{(\qe,\delta)}=I(X_t;Y_t|X_{t-1})\leq \I(X;Y)$, where the inequality is generically strict if the environmental process is not i.i.d.. The value of acquired information, $I_\pe^{(\qe,\delta)}$, is thus {\it not} given by the rate of mutual information $\I(X;Y)$, except in special cases such as when no correlations are present between successive environmental states (i.i.d.~environment).\\

$I_\pe^{(\qe,\delta)}$ does not indeed correspond to the rate of mutual information, but to the rate of directed information~\cite{Marko73,Massey90,PermuterKim08}, generally defined by
\beq\label{eq:crDI}
I(Y^T\to X^T)\equiv \sum_{t=1}^T I(X_t;Y^t|X^{t-1}).
\eeq
For a Markov environmental process, conditioning with respect to $X^{t-1}$ is equivalent to conditioning with respect to $X_{t-1}$ and $I(X_t;Y^t|X_{t-1})=I(X_t;Y_t|X_{t-1})$, so that the generic term of the sum equates the conditional mutual information obtained in Eq.~\eqref{eq:condMI}. If $\I(Y\to X)\equiv \lim_{T\to\infty}I(Y^T\to X^T)/T$ denotes the rate of directed information, we have therefore $I_\pe^{(\qe,\delta)}=\I(Y\to X)$. To understand the origin of the difference between $\I(Y\to X)$ and the rate of mutual information $\I(X;Y)$, we may similarly expand the mutual information $I(X^T;Y^T)$ using the chain rule~\cite{CoverThomas91}:
\beq\label{eq:crMI}
I(X^T;Y^T)=\sum_{t=1}^T I(X_t;Y^T|X^{t-1}).
\eeq
In this expression, $Y^t$ in Eq.~\eqref{eq:crDI} is replaced by $Y^T=(Y^t,Y_{t+1},\dots,Y_T)$. Consequently, $I(Y^T\to X^T)\leq I(X^T;Y^T)$ and the difference may be interpreted as the information that would be gained about the current environmental state $x_t$ from knowing the future signals $y_{t+1},\dots,y_T$; these signals are indeed informative about $x_t$, since they are correlated to $x_t$ through $x_{t+1},\dots,x_T$, although they are not accessible at time $t$ for a strategy $\pi(\s|x_{t-1},y_t)$ which relies only on the current signal $y_t$ (keeping memory of the past signals $y^{t-1}$ does not make a difference in the present context where $x_{t-1}$ is available). The mutual information $I(X^T;Y^T)$ thus accounts for all statistical correlations between $X^T$ and $Y^T$, while the directed information $I(Y^T\to X^T)$ accounts only for the correlations that are consistent with the constraints of causality imposed on $\pi(\s|x_{t-1},y_t)$. Consistently with this interpretation, the difference $I(X^T;Y^T)-I(Y^T\to X^T)$ can be shown to be $I(X^{T-1}\to Y^T)$.\\

{\bf Cost of uncertainty in presence of acquired information --} Similarly, the uncertainty cost in presence of side-information, given in Eq.~\eqref{eq:noH1}, does not correspond to the rate $\H(X|Y)$ of the conditional entropy $H(X^T|Y^T)$, but instead to the rate $\H(X\|Y)$ of the causally conditional entropy $H(X^T\|Y^T)$~\cite{Kramer98,PermuterKim08}, which is generally defined by
\beq
H(X^T\|Y^T)\equiv\sum_{t=1}^T H(X_t|X^{t-1},Y^t).
\eeq
For comparison, $H(X^T|Y^T)$ can be similarly expressed with $Y^t$ replaced by $Y^T$ in each term of the sum, thus showing that $H(X^T\|Y^T)\geq H(X^T|Y^T)$. In the context of our model, $H(X_t|X^{t-1},Y^t)=H(X_t|X_{t-1},Y_t)$, and hence Eq.~\eqref{eq:noH1} indicates that $H_\pe^{(\qe,\delta)}=\H(X\|Y)$. As the conditional entropy is related to the mutual information by $I(X^T;Y^T)=H(X^T)-H(X^T|Y^T)$, the causally conditional entropy is related to the directed information by $I(Y^T\to X^T)=H(X^T)-H(X^T\|Y^T)$, or, in terms of rates, $\I(Y\to X)=\H(X)-\H(X\|Y)$.\\

{\bf Conclusion --} The conclusions that the uncertainty cost is given by the rate of a causally conditional entropy, which is greater than the rate of a conditional entropy,
\beq\label{eq:Heq1}
H_\pe^{(\qe,\delta)}=\H(X\|Y)\geq \H(X|Y),
\eeq
and the value of acquired information by the rate of directed information, which is smaller than the rate of mutual information,
\beq\label{eq:Ieq1}
I_\pe^{(\qe,\delta)}= \I(Y\to X)\ \leq\ \I(X;Y),
\eeq
can be extended beyond Markov processes to more general ergodic stochastic processes, provided one allows for arbitrary long memory, i.e., strategies of the form $\pi(x_t|x^{t-1},y^t)$~\cite{PermuterKim08}. More generally, the notion of directed information appears as the relevant generalization of the notion of mutual information when causal relations, and not merely statistical relations, must be taken into account~\cite{Marko73,Massey90}; for instance, while the capacity of memoryless channels is expressed in terms of a mutual information, the capacity of channels with feedback involves a directed information~\cite{Kim08}.\\

Coming back to our model, in absence of the simplifying assumption (A3) that the multiplication matrix $f(\s;x)$ is diagonal, problems involving both acquired and inherited information are generally difficult to solve; in particular, no closed-form expression for the growth rate generalizing Eq.~\eqref{eq:closedformed} is available. Horse race models are an exception, due to the fact that the history of past types of any individual mirrors the history of past environmental states, since only individuals with $\s_t=x_t$ survived at time $t$. This reduces the problem to an effectively feedforward problem, where $\pi(\s_t|\s_{t-1},y_t)$ does not actually depend on the "control variable" $\s_{t-1}$, but only on the "primary variables" $x_{t-1}$ and $y_t$. An other solvable case, for essentially the same reason, is the limit where any given environmental state lasts long enough for a single type to dominate the population~\cite{KussellLeibler05}: we show in  appendix~\ref{sec:timing} how the problems of delay and timing that generally arise in correlated environments with inherited information can be treated in this case.\\ 

\section{Individual stochasticity and distributed information}\label{sec:indiv}

{\bf Cost of uncertainty --} Retaining the assumptions (A1) and (A3) but now relaxing (A2) by allowing each individual to perceive a different signal from the environment leads to a different generalization of the definitions of entropy and mutual information, with no equivalent in the context of models of financial investment. In this case, the expression for the growth rate, Eq.~\eqref{eq:closedformed}, is
\beq
\Lambda_{\pe;f}^{(\qe,\qi)}(\pi)=\sum_{x,x'}\qe(x'|x)\pes(x)\ln\left(\sum_{y}f(x)\pi(x|y)\qi(y|x')\right).
\eeq
Following the derivation given in Sec.~\ref{sec:horseraces}, its optimal value can again be decomposed in two terms,
\beq
\hat \Lambda_{\pe;f}^{(\qe,\qi)}=\hat \Lambda_{\pe;f}^{(\delta,\delta)}-H_\pe^{(\qe,\qi)}.
\eeq
The second term, which is again independent of $f$,
\beq
H_\pe^{(\qe,\qi)}\equiv\min_\pi \sum_{x,x'}\qe(x'|x)\pes(x)\ln\left(\sum_y\pi(x|y)\qi(y|x')\right)^{-1},
\eeq
generalizes the notions of entropy $H_\pe^{(\eta,\delta)}=H(X_t)$ and conditional entropy $H_\pe^{(\qe,\delta)}=H(X_t|Y_t)$ obtained for horse races in Sec.~\ref{sec:horseraces}~\footnote{$H_\pe^{(\qe,\qi)}$ can also be written
$$H_\pe^{(\qe,\qi)}=H(X_t|X'_t)+\min_\pi\E_{X'_t}\left[D\left(\P_{X_t|X_t'}(x|X_t')\| \pi*\qi(x|X'_t)\right)\right],$$
where $\pi*\qi(x|x')=\sum_y\pi(x|y)\qi(y|x')$ and where $\P_{X_t|X'_t}(x|x')$ is the optimal strategy for the same problem where $\qi=\delta$, i.e., $\P_{X_t|X_t'}(x|x')=\qe(x'|x)\pe(x)/(\sum_{z}\qe(x'|z)\pe(z))$.}. From the concavity of the logarithm (Jensen's inequality), 
\beq\label{eq:ineqH}
H(X_t|X'_t)\leq H_\pe^{(\qe,\qi)}\leq H(X_t|Y_t),
\eeq
where, following the usual notations, $X_t'$ refers to the random variable for the component $x'_t$ of the signal defined at the population level, and $Y_t$ to the random variable for the signal $y_t$ effectively perceived by an individual (see Fig.~\ref{fig:scheme}).\\

\begin{figure}
\includegraphics[width=.4\linewidth]{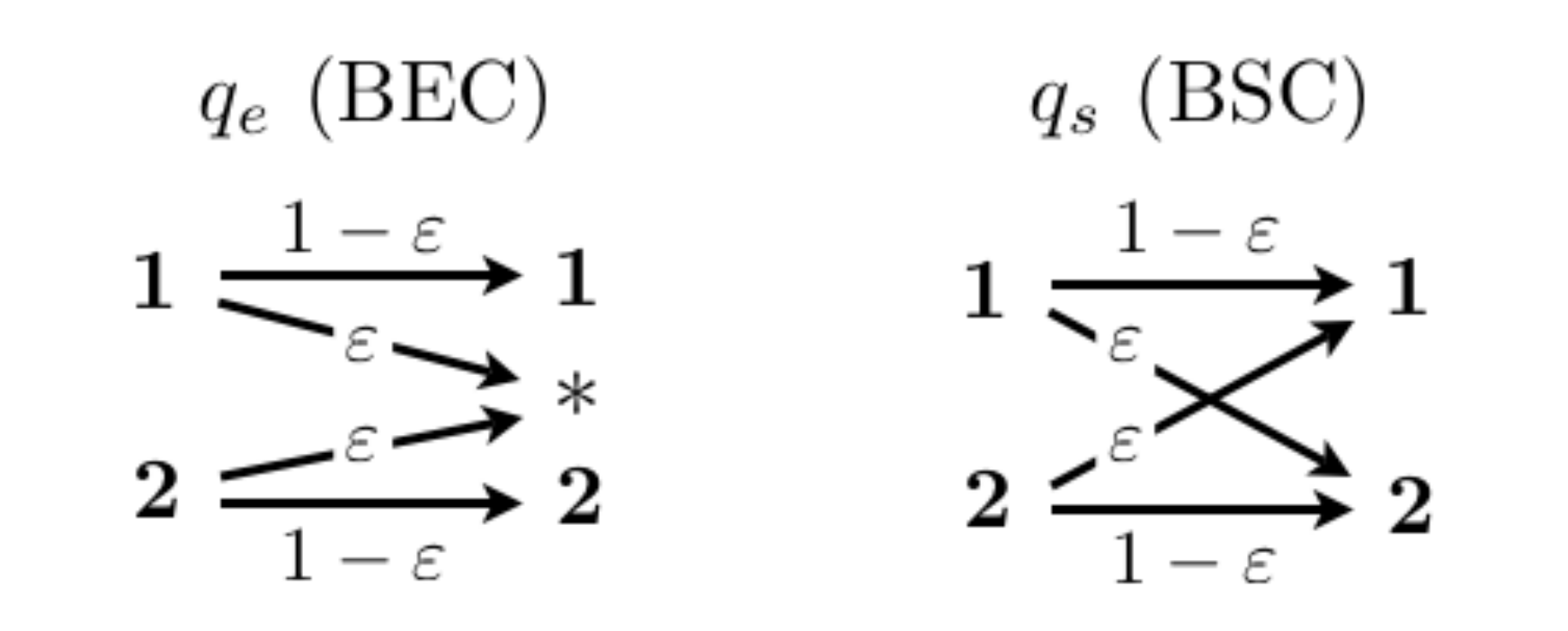}
\caption{Models of communication channels for $\qe$ and $\qi$ when the environment alternates between two states 1 and 2. Left: the binary erasure channel $q_e$ (BEC), which erases the input with probability $\varepsilon$ and transmits it faithfully with probability $1-\varepsilon$. Right: the binary symmetric channel $q_s$ (BSC),  which exchanges the input with probability $\varepsilon$ and transmits it unchanged with probability $1-\varepsilon$. The channels $\delta$ and $\eta$ presented in Fig.~\ref{fig:trivialchannels} represent extreme cases: the noiseless channel $\delta$ corresponds to the BEC or the BSC with $\varepsilon=0$, and the informationless channel $\eta$ to the BEC with $\varepsilon=1$, or the BSC with $\varepsilon=1/2$.}\label{fig:channels}
\end{figure}

\begin{figure}
\begin{minipage}[c]{.46\linewidth}
\centering
\includegraphics[width=.975\linewidth]{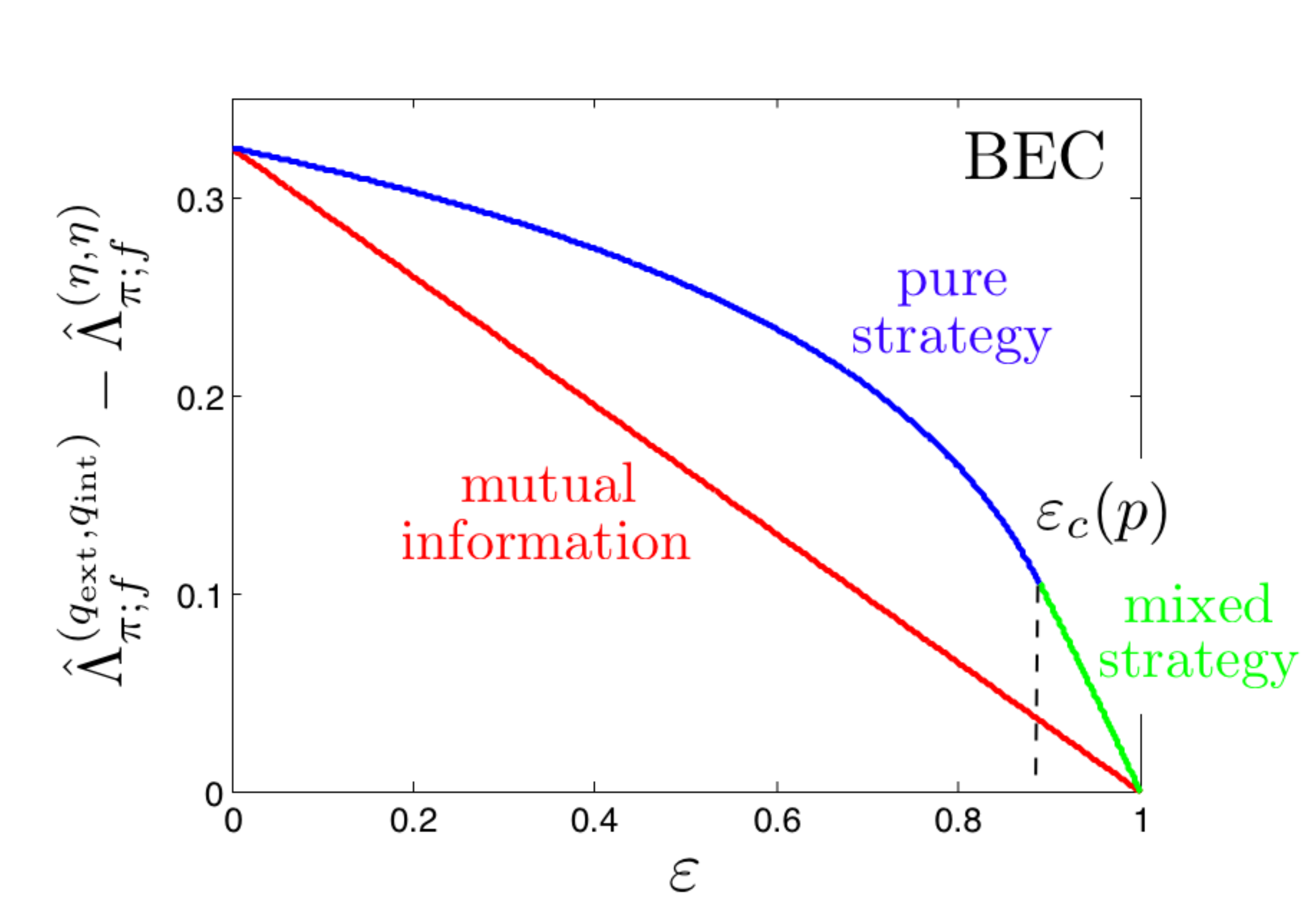}
\caption{\small Value of information for a two-state model with information transmitted through a binary erasure channel situated either at the population level (red curve) or at the individual level (blue and green curve). The probability of the environmental state 1 is here fixed to $\pe(1)=0.1$. The binary erasure channel $q_e$, defined in Fig.~\ref{fig:channels}, has a probability of erasure $\varepsilon$ which is varied along the $x$-axis. When the channel is at the population level, the value of information $\hat \Lambda_{\pe;f}^{(q_e,\delta)}-\hat \Lambda_{\pe;f}^{(\eta,\eta)}$ corresponds to the mutual information between the input and output signals which, for the binary erasure channel, is a linear function of $\e$ (red line). When the same channel is at the individual level, the value of information $\hat \Lambda_{\pe;f}^{(\delta,q_e)}-\hat \Lambda_{\pe;f}^{(\eta,\eta)}$ is generally higher (blue and green curve), and a transition occurs at $\varepsilon_c(\pe)=(1-2\pe(1))/(1-\pe(1))=0.88$: for $\varepsilon<\varepsilon_c(\pe)$, the optimal strategy is a pure strategy with $\hat \pi(1|*)=0$ (blue part), while for $\varepsilon>\varepsilon_c(\pe)$, it becomes a mixed strategy with both $\hat \pi(1|*)>0$  and $\hat \pi(2|*)>0$ (green part). The calculations are detailed in appendix~\ref{app:bec}.\label{fig:BEC}}
\end{minipage} \hfill
\begin{minipage}[c]{.46\linewidth}
\centering
\includegraphics[width=.975\linewidth]{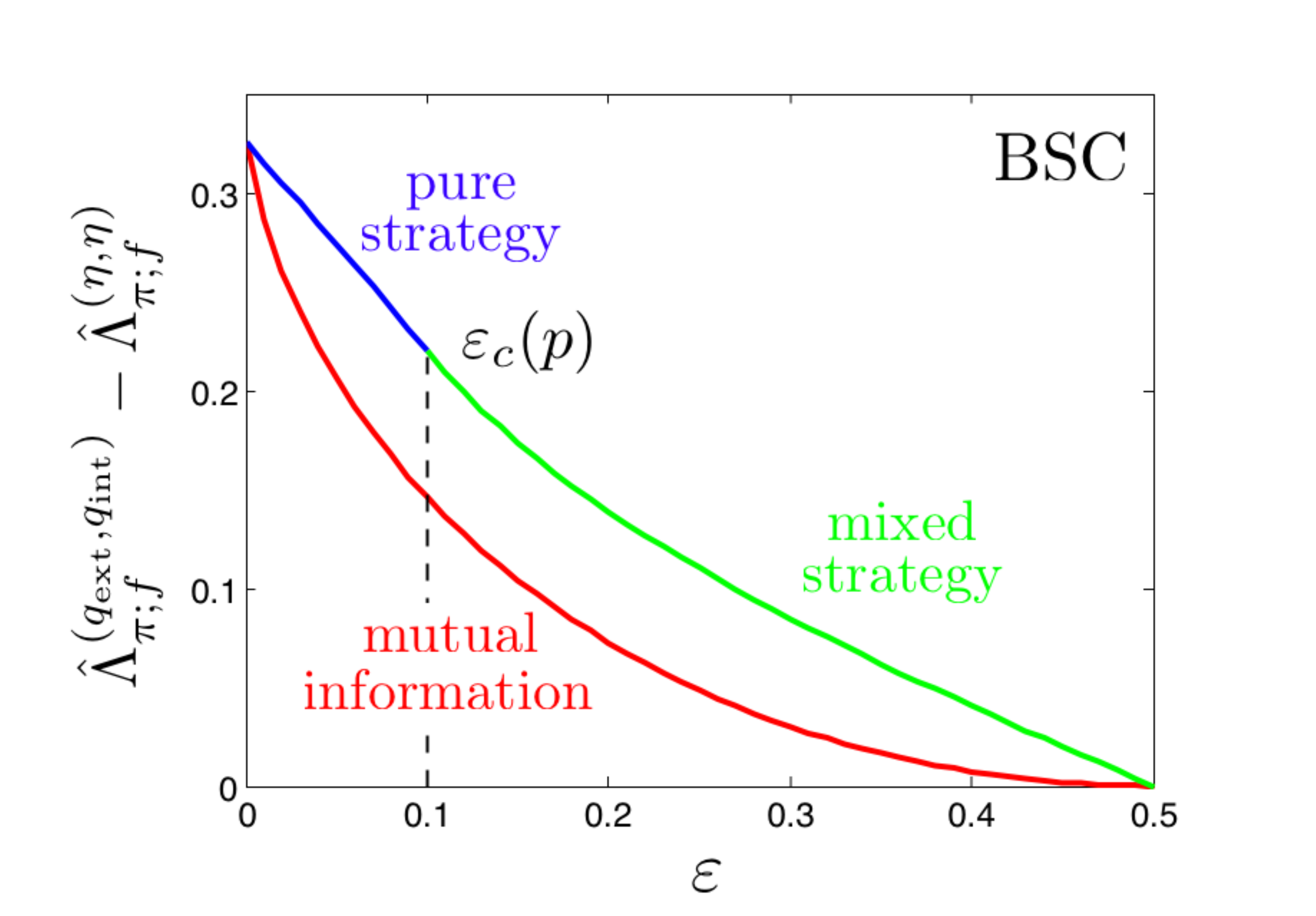}
\caption{\small Value of information for a two-state model with information transmitted through a binary symmetric channel situated either at the population level (red curve) or at the individual level (blue and green curve). The probability of the environmental state 1 is here fixed to $\pe(1)=0.1$. The binary symmetric channel $q_s$, defined in Fig.~\ref{fig:channels}, has a probability of error  $\varepsilon$ which is varied along the $x$-axis. When the channel is at the population level, the value of information $\hat \Lambda_{\pe;f}^{(q_s,\delta)}-\hat \Lambda_{\pe;f}^{(\eta,\eta)}$ corresponds to the mutual information between the input and output signals (red curve). When the same channel is at the individual level, the value of information $\hat \Lambda_{\pe;f}^{(\delta,q_s)}-\hat \Lambda_{\pe;f}^{(\eta,\eta)}$ is generally higher (blue and green curve), and a transition occurs at $\varepsilon_c(\pe)=\pe(1)=0.1$: for $\varepsilon<\varepsilon_c(\pe)$, the optimal strategy is to adopt a pure strategy when receiving either of the two possible signals 1 and 2, i.e., $\hat \pi(1|1)=\hat \pi(2|2)=1$ (blue part), while for $\varepsilon>\varepsilon_c(\pe)$ the optimal strategy is a mixed strategy with $\hat \pi(2|2)=1$ but $0<\hat \pi(1|1)<1$ (green part). The calculations are detailed in appendix~\ref{app:bsc}.\label{fig:BSC}}
\end{minipage}
\end{figure}

As an illustration of the properties of this generalized entropy, showing in particular that, generically, $H_\pe^{(\qe,\qi)}<H(X_t|Y_t)$, we compare in Fig.~\ref{fig:BEC} and~\ref{fig:BSC} the benefits of the same channel $q$ located either at the population level, $\qe=q$, $\qi=\delta$, or at the individual level, $\qe=\delta$, $\qi=q$, taking for $q$ two classical examples of communication channels defined in Fig.~\ref{fig:channels} (the details of the calculations are presented in appendix~\ref{app:solvable}).\\

{\bf Value of information --} The fact apparent in Fig.~\ref{fig:BEC}-\ref{fig:BSC} that the same communication channel $q$ induces less uncertainty when located at an individual level than at a population level, i.e., $\hat \Lambda_{\pe;f}^{(q,\delta)}\leq \hat \Lambda_{\pe;f}^{(\delta,q)}$, holds generally, again as a consequence of Jensen's inequality,
\beq\label{eq:leq}
\hat \Lambda_{\pe;f}^{(\qi*\qe,\delta)}\leq\hat \Lambda_{\pe;f}^{(\qe,\qi)}\leq \hat \Lambda_{\pe;f}^{(\delta,\qi*\qe)},
\eeq
where $\qi*\qe$ denotes the convolution of $\qi$ and $\qe$, i.e., $\qi*\qe(y|x)=\sum_{x'}\qi(y|x')\qe(x'|x)$. An important implication is that the mutual information between the source $X_t$ and the perceived signal $Y_t$ does {\it not} represent an upper bound for the value of acquired information. From the relation $H_\pe^{(\qe,\qi)}+I_\pe^{(\qe,\qi)} = \hat\Lambda_{\pe;f}^{(\delta,\delta)}- \hat\Lambda_{\pe;f}^{(\eta,\eta)}$, we have indeed a relation dual to Eq.~\eqref{eq:ineqH} for the value of information $I_\pe^{(\qe,\qi)}\equiv \hat\Lambda_\pe^{(\eta,\eta)}-\hat\Lambda_\pe^{(\qe,\qi)}$:
\beq
I(X_t;Y_t)\leq I_\pe^{(\qe,\qi)}\leq I(X_t;X'_t).
\eeq
Informally, we may say that the value of the information acquired collectively by the population exceeds the value of the information acquired by any of its members. This result contrasts with the law of requisite variety derived in other contexts which states that the mutual information $I(X_t;Y_t)$ between the environmental fluctuation $X_t$ and the signal $Y_t$ derived from it sets an upper limit on the value of information for control~\cite{Ashby58,TouchetteLLoyd00}. In comparison with the mutual information $I(X_t;Y_t)$, $I_\pe^{(\qe,\qi)}$ is not symmetrical in $X_t$ and $Y_t$, although it similarly satisfies $I_\pe^{(\qe,\qi)}=0$ if and only if $X_t$ and $Y_t$ are independent.\\

{\bf Optimal strategy and Bayesian inference --} Another remarkable feature displayed in Fig.~\ref{fig:BEC} and~\ref{fig:BSC} is the possible existence of a critical level of noise $\varepsilon_c(\pe)$ below which a stochastic response is not required for achieving optimal growth. This contrasts with the horse race model, where a non-deterministic response is required not only to achieve an optimal growth, but even more fundamentally to avoid extinction. Here, the diversification of the population caused by the deterministic response of individuals perceiving stochastic signals is optimal at low error rates. Although estimation and decision can be separated in principle ~\cite{Witsenhausen71}, and although a Bayesian computation, as in Eq.~\eqref{eq:stratB}, would provide an optimal estimation, the simplest implementation of the optimal strategy involves here no computation at all:  when $\varepsilon<\varepsilon_c(\pe)$, the individual can process the signal as if it were perfectly reliable~\footnote{The optimal strategy does not require a Bayesian computation, but it nevertheless follows a Bayesian logics, in the sense of the word given in stochastic adaptive control theory~\cite{Berger85}.}. This situation is analogous to the situation with optimal source-channel communication: although in principle a solution can always be obtained by treating separately the problems of source compression and channel coding~\cite{Shannon48}, a computationally much simpler solution may be available, which in some cases does not involve any coding at all~\cite{Gastpar03}. Living systems are unlikely to solve stochastic control problems by relying on the estimation-decision separation principle, as they are unlikely to solve communication problems by relying on the source-channel separation principle~\cite{Berger03}.

\begin{figure}
\begin{minipage}[c]{.46\linewidth}
\centering
\includegraphics[width=.8\linewidth]{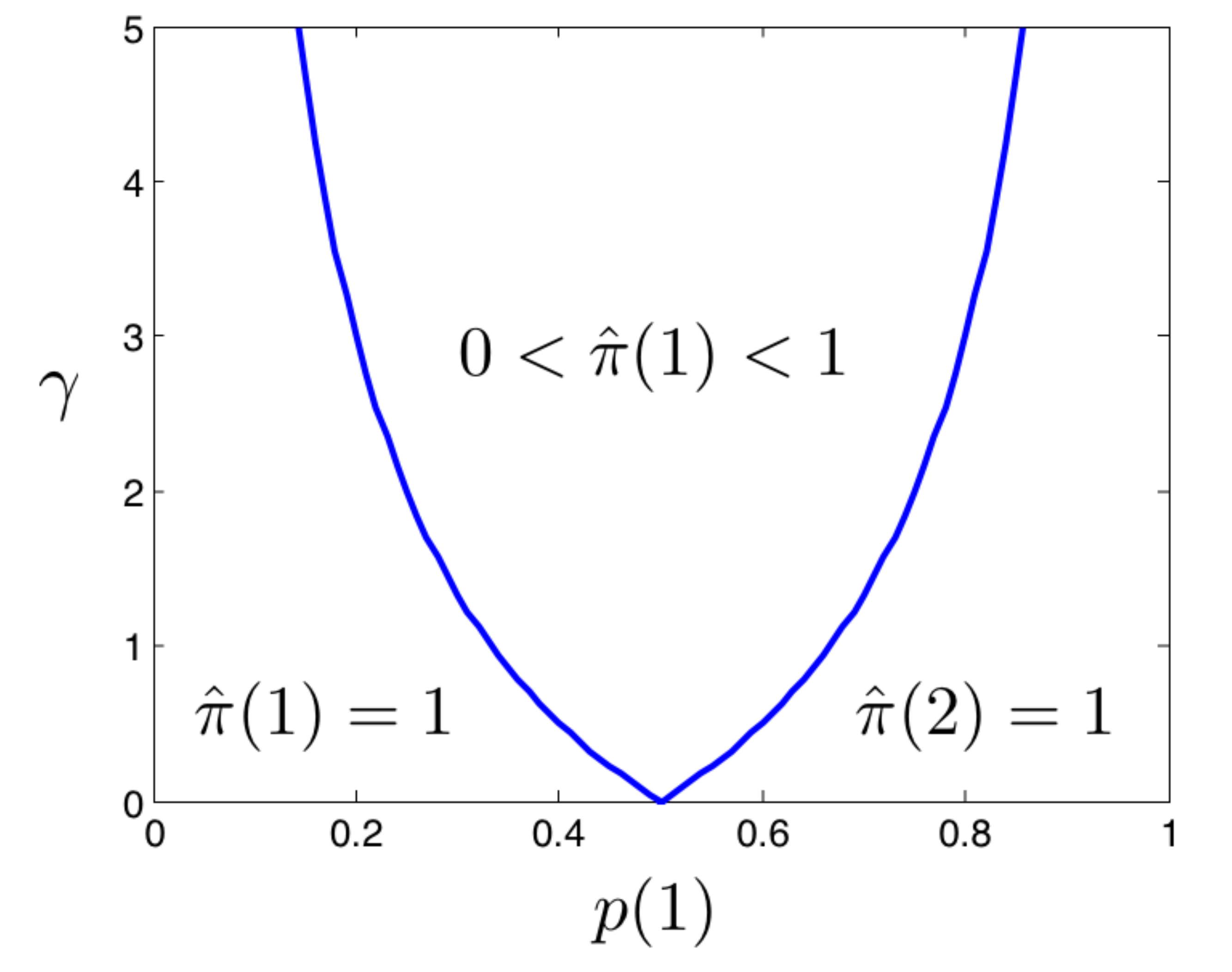}
\caption{\small Nature of the optimal strategy for a model with 2 types and 2 environmental states whose probabilities $\pe(1)$ and $\pe(2)=1-\pe(1)$ are varied along the $x$-axis, and with multiplication rates whose ratio $\gamma=f(1;1)/f(2;1)-1=f(2;2)/f(1;2)-1$ is varied along the $y$-axis (this ratio may be thought as quantifying the dissimilarity between the two types). No information is assumed to be available. The blue curves delineate the regions of the parameter space where the optimal strategy involves switching, with $0<\hat \pi(1)<1$, from the regions where one of the two types is excluded from the optimal strategy, corresponding to homogeneous populations, with $\hat \pi(1)=1$ or $\hat \pi(2)=1$. The location of the transitions is given by $\g_c^{(1)}=(1-\pe(1))/\pe(1)-1$ and $\g_c^{(2)}=\pe(1)/(1-\pe(1))-1$. The calculations are detailed in appendix~\ref{app:2state}.
\ \\
\ \\
\ \\
\ \label{fig:quant}}
\end{minipage} \hfill
\begin{minipage}[c]{.46\linewidth}
\centering
\includegraphics[width=.95\linewidth]{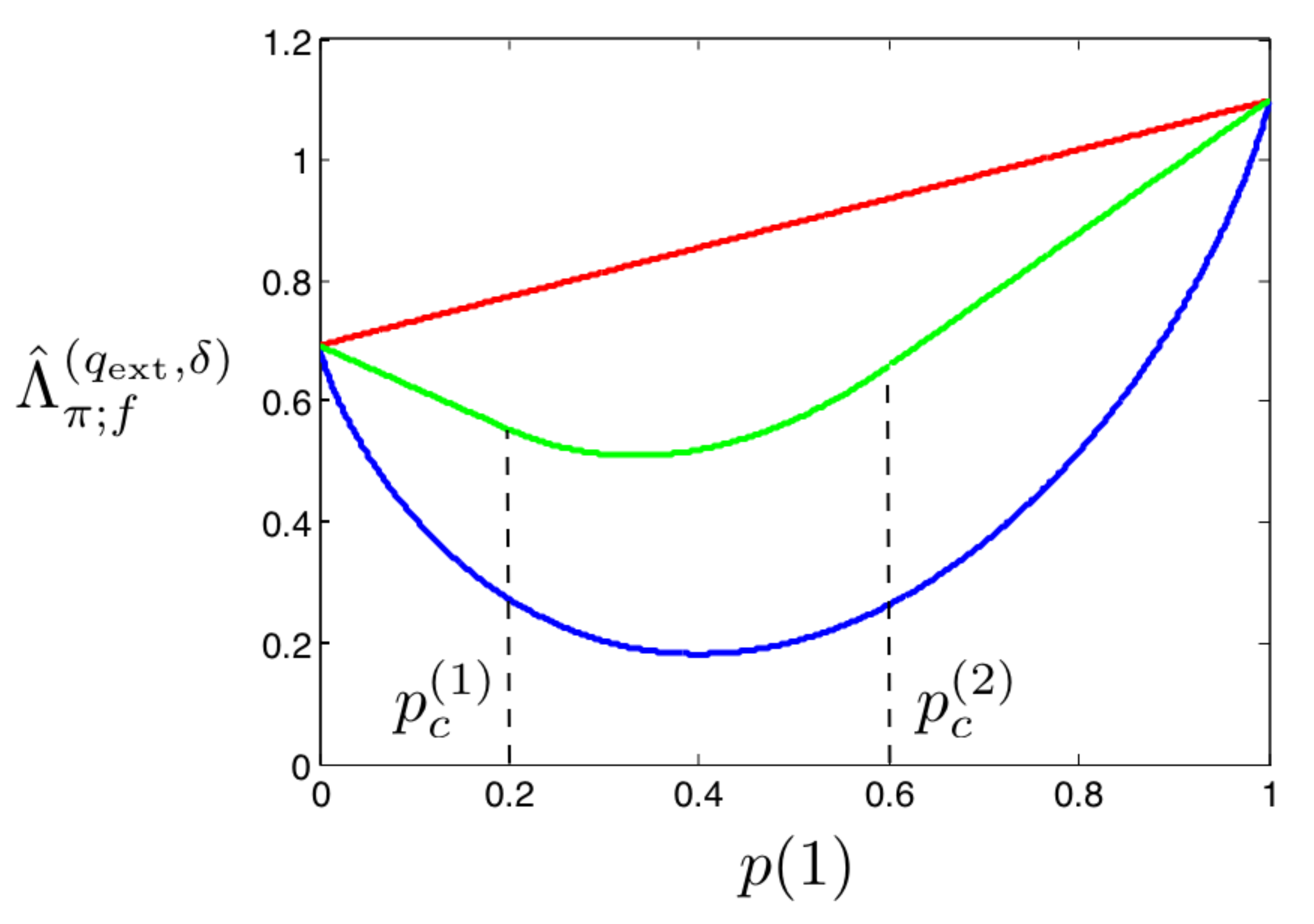}
\caption{\small In green: Optimal growth rate $\hat\Lambda^{(\eta,\eta)}_\pe$ for a model in absence of information with 2 types and 2 environmental states, as a function of the probability $\pe(1)$ of the first environmental state. The values of multiplication rates are $f(1;1)=3$, $f(2;2)=2$, $f(2;1)=2$,  and $f(1;2)=1$. The dashed lines represent transitions between mixed and pure strategies: for $\pe(1)<\pe_c^{(1)}=0.2$ the type $\s=1$ is excluded from the optimal strategy, while for $\pe(1)>\pe_c^{(2)}=0.6$ this is the case for $\s=2$. In red: Optimal growth rate $\hat\Lambda_{\pe;f}^{(\delta,\delta)}$ for the same model in presence of complete information. In blue: Optimal growth rate $\hat\Lambda_{\pe,f}^{(\delta,\delta)}-H(X_t)$ for a corresponding horse race model with $f(1;1)=3$, $f(2;2)=2$ and $f(2;1)=f(1;2)=0$ and no information. This example shows that the uncertainty cost $\hat\Lambda_\pe^{(\delta,\delta)}-\hat\Lambda^{(\eta,\eta)}_\pe$ (difference between the red and green curves) is generally strictly smaller than the entropy of the environment $H(X_t)$ (difference between the red and blue curves). The calculations are detailed in appendix~\ref{app:2state}.\label{fig:lambda}}
\end{minipage}
\end{figure}

\section{General multiplication rates and functional information}\label{sec:f}

Retaining the assumptions (A1) and (A2) but relaxing (A3) leads to a different departure from the usual concepts of communication theory. Now the cost of uncertainty $\hat \Lambda_{\pe;f}^{(\delta,\delta)}-\hat \Lambda_{\pe;f}^{(\qe,\qi)}$ and the value of acquired information $\hat \Lambda_{\pe;f}^{(\qe,\qi)}-\hat \Lambda_{\pe;f}^{(\eta,\eta)}$ are no longer necessarily independent of the multiplication rates $f(\s;x)$, and cannot therefore be written as statistical quantities $H_\pe^{(\qe,\qi)}$ or $I_\pe^{(\qe,\qi)}$ depending only on the transition matrices $\pe(x_t|x_{t-1})$, $\qe(x'_t|x_t)$ and $\qi(y_t|x'_t)$.\\

{\bf Uncertainty cost --} A very special feature of models satisfying assumption (A3), i.e., models where the multiplication rates have a diagonal form, with $f(\s;x)=f(x)$ if $\s=x$ and 0 otherwise, is that the environmental states $x$ and the individual types $\s$ are in one-to-one correspondence. The environment is however generally defined independently of any reference to the internal states of the individuals of the population. We should therefore not expect a quantity like the entropy rate of the environmental process, $\H(X)$, to correctly capture the cost of uncertainty, which depends essentially on the definition of the internal states of the individuals. The environmental states may indeed specify details that are irrelevant to the growth of the population, say the positions of distant stars, which inflate arbitrarily the entropy rate $\H(X)$ without influencing the uncertainty cost $\hat \Lambda_{\pe;f}^{(\delta,\delta)}-\hat \Lambda_{\pe;f}^{(\qe,\qi)}$. As an example, consider an horse race where each distinct environmental state corresponds to a distinct ordered list of arrival of all the horses participating to the race; this description indeed includes useless information if only the first horse has a non-zero pay-off. In such a case, we may still capture the uncertainty cost by a statistical quantity by partitioning the environmental states into exclusive sets $\X(\s)$ grouping the lists where horse $\s$ is first, such that $f(\s;x)=f(\s)$ if $x\in\X(s)$ and 0 otherwise: assuming i.i.d.~races, the uncertainty cost then correspond to the entropy of the coarse-grained description $H(\X_t)$ rather than the entropy $H(X_t)$, with obviously $H(X_t)\geq H(\X_t)$ (see also appendix~\ref{SI:Hbound}). More generally, the uncertainty cost ignores any stochastic element of the environment that is irrelevant for the growth of the population, but nevertheless contributes to the entropy rate $\H(X)$.\\

In addition, when several types $\s$ have non-zero multiplication rates $f(\s;x)>0$ in a given environmental state $x$, the non-optimal but yet surviving types contribute to the growth although they are not associated with the exact prediction of the optimal type $\s$ for the given environment $x$. Again, this implies that an entropic measure based only on the environmental process tends to overestimate the uncertainty cost. We show in appendix~\ref{SI:Hbound} that the following bound holds: 
\beq\label{eq:Hbound}
\hat \Lambda_{\pe;f}^{(\delta,\delta)}-\hat\Lambda_{\pe;f}^{(\qe,\qi)}\leq  H_\pe^{(\qe,\qi)}.
\eeq
Here, the generalized entropy $H_\pe^{(\qe,\qi)}$ is the uncertainty cost for a horse race model with same channels $\pe(x_t|x_{t-1})$, $\qe(x'_t|x_t)$ and $\qi(y_t|x'_t)$. As defined in the previous sections, this generalized entropy is independent of the value of the multiplication rates $f$. The quantity $\hat \Lambda_{\pe;f}^{(\delta,\delta)}-\hat\Lambda_{\pe;f}^{(\qe,\qi)}$ can be seen as a measure of uncertainty that refines $H_\pe^{(\qe,\qi)}$ by accounting for the effective reduction of uncertainty due to the redundancy between environments and types encoded in $f$. This is a further refinement over the concept of entropy $H_\pe^{(\qe,\qi)}$, which itself can be seen as refining Hartley's measure $\ln n$ by accounting for the effective reduction of uncertainty due to the unequal probabilities of the different environmental states. In the two cases, the refinement takes the form of an inequality, with equality if $f$ is diagonal in the first case, and if all the environmental states are equiprobable in the second case.\\

{\bf Value of information --} The corresponding inequality holds for the value of information, with
\beq\label{eq:ineqI}
\hat \Lambda_{\pe;f}^{(\qe,\delta)}-\hat \Lambda_{\pe;f}^{(\eta,\delta)}\leq I_\pe^{(\qe,\delta)}.
\eeq
In particular, under the assumptions (A1) and (A2) such that $I_\pe^{(\qe,\delta)}$ is given by the mutual information $I(X_t,Y_t)$, the value of information is bounded by $I(X_t,Y_t)$. The deviation of $\hat \Lambda_{\pe;f}^{(\qe,\delta)}-\hat \Lambda_{\pe;f}^{(\eta,\delta)}$ from $I_\pe^{(\qe,\delta)}$, when the multiplication rates are non-diagonal, can be interpreted as arising from the fact that the environmental states have no longer an exclusive "meaning", in the sense that the same environment can be beneficial to different types, and different environments to the same type.
A noticeable feature of models with non-diagonal multiplication matrices is also that the optimal strategy may actually exclude some types $\s$, i.e., we may have $\hat \pi(\s|y)=0$ for some $\s$. A trivial example is when two types $\s$ and $\s'$ are present, for which $f(\s;x)> f(\s';x)$ in any environmental state $x$, in which case the optimal strategy will never populate $\s'$. A less trivial, yet analytically solvable class of models which display the same feature is defined by extending (A3) to the case where the off-diagonal terms of the matrix $f(\s;x)$ are non-zero but constant, i.e., $f(\s;x)=f(x)$ for $\s=x$ and $f(\s;x)=\tilde f(x)<f(x)$ for $\s\neq x$  (see appendix~\ref{app:solvable}); in particular for $n=2$ states, the model is solvable for arbitrary matrices $f$, as illustrated in Fig.~\ref{fig:quant} and~\ref{fig:lambda}~\footnote{Another interesting subclass of models is when $f(\s;x)=g(x)\phi(x|\s)$ where $\phi$ is a transition matrix, i.e., $\sum_x\phi(x|\s)=1$; the growth rate can then be written $\Lambda_{\pe;f}^{(\eta,\delta)}=\E_X[\ln g(X)]-H(X)-D(\pe\|\phi*\pi)$ and the optimal strategy is given by the minimization of $D(\pe\|\phi*\pi)$. The problem to be solved to find the optimal strategy $\hat \pi$ is then equivalent to a problem of blind source identification, i.e., the problem of inferring the source $\pi$ of the inputs of a communication channel $\phi$ given the distribution $\pe$ of its outputs.}. Another important solvable class of models is when a separation of time scales allows for an adiabatic approximation, as presented in appendix~\ref{sec:timing}.\\

\begin{table}
\renewcommand{\arraystretch}{1.6} 
\setlength{\tabcolsep}{5pt} 
 \begin{center}
 \begin{tabular}{r|c|c|c||c} 
Assumptions\hspace{.3cm}  & Extra assumption &\hspace{.3cm} Value of information\hspace{.3cm} & Cost of uncertainty & Sec. \\ 
\hline \hline
& $\_$ & & & \\
 &  & $\hat \Lambda_{\pe;f}^{(\qe,\qi)}-\hat \Lambda_{\pe;f}^{(\eta,\eta)}$ &  $\hat \Lambda_{\pe;f}^{(\delta,\delta)}-\hat \Lambda_{\pe;f}^{(\qe,\qi)}$ & VII\\
\hline
(A3)\hspace{.3cm} & no survival for $\s_t\neq x_t$ &  &  &  \\
 & $f(\s_t;x_t)=f(x_t)\delta(x_t|\s_t)$ & $I_p^{(\qe,\qi)}$ & $H_p^{(\qe,\qi)}$ & VI\\
\hline
(A2)\ (A3)\hspace{.3cm} & no individuality & directed information& causally conditional entropy & \\
 & $\qi(y_t|x'_t)=\delta(y_t|x'_t)$ & $\I(Y\to X)$ & $\H(X\|Y)$ & V\\
\hline
(A1)\ (A2)\ (A3)\hspace{.3cm} & no feedback & mutual information  & conditional entropy & \\
 & $\pi(\s_t|\s_{t-1},y_t)=\pi(\s_t|y_t)$ & $I(X_t;Y_t)$ & $H(X_t|Y_t)$ & IV\\
 \end{tabular}
 \end{center}
 \caption{Expressions for the value of information and cost of uncertainty under different assumptions. The top row corresponds to the most general model and each subsequent row involves an additional assumption, indicated in the second column. The last row thus defines the most restrictive model, which is the horse model from which we started in Sec.~\ref{sec:horseraces}. We then presented the implications of relaxing successively the various assumptions that it involves, thus moving up in this table. The different measures of information are related by $\hat \Lambda_{\pe;f}^{(\qe,\qi)}-\hat \Lambda_{\pe;f}^{(\eta,\eta)}\leq I_\pe^{(\qe,\qi)}\leq \I(X'\to X)\leq I(X_t;X'_t)$ and  $\I(Y\to X)\leq I_\pe^{(\qe,\qi)}$, where $\I(X'\to X)=I_\pe^{(\qe,\delta)}$ and $\I(Y\to X)=I_\pe^{(\qe*\qi,\delta)}$: see Eqs.~\eqref{eq:Ieq1}, \eqref{eq:leq}, \eqref{eq:ineqI}. Similarly, the different measures of uncertainty are related by $\hat \Lambda_{\pe;f}^{(\delta,\delta)}-\hat \Lambda_{\pe;f}^{(\qe,\qi)}\leq H_\pe^{(\qe,\qi)}\leq H(X_t|Y_t)\leq \H(X\| Y)$ and  $\H(X\| X')\leq H_\pe^{(\qe,\qi)}$, where $\H(X\| X')=H_\pe^{(\qe,\delta)}$ and $\H(X\| Y)=H_\pe^{(\qe*\qi,\delta)}$: see Eqs.~\eqref{eq:Heq1}, \eqref{eq:ineqH}, \eqref{eq:Hbound}.}\label{fig:tablesum}
\end{table}

{\bf General conclusion --} To sum up the results of the last three sections, the relaxations of the assumptions (A1), (A2) and (A3) lead to generalizations of the notions of entropy and mutual information in three different directions: (i) to account for the constraints of causality (Sec.~\ref{sec:dirinfo}); (ii) to account for the level at which information is processed (Sec.~\ref{sec:indiv}); (iii) to account for the meaning of information encoded in the matrix $f(\s;x)$ (this section). In the cases (i) and (iii), which had been previously studied from the standpoint of financial investment, the mutual information appears as an upper limit for the value of acquired and inherited information, consistently with Ashby's law of requisite variety~\cite{Ashby58}; this limit cannot generally be reached, and a tighter and achievable upper bound is provided by $I_\pe^{(\qe,\qi)}$. In the case (ii), which is specific to the biological interpretation of the model, the fundamental limit $I_\pe^{(\qe,\qi)}$ can be greater than the mutual information $I(X_t;Y_t)$. In general, all three assumptions (A1), (A2) and (A3) may be jointly violated, and the uncertainty cost and value of information need to be measured accordingly. These conclusions are summarized in Table~\ref{fig:tablesum}. The problem of measuring the degree of adaptation of a population with given communication channels $\qe$ and $\qi$ can be treated as well. As shown in appendix~\ref{SI:Dbound}, the identity involving the relative entropy that emerged from the analysis of horse race models in Sec.~\ref{sec:horseraces}, Eq.~\eqref{eq:D}, is more generally replaced by an inequality.

\section{Generalizations}\label{sec:general}

\begin{figure}
\begin{minipage}[c]{.36\linewidth}
\centering
\includegraphics[width=\linewidth]{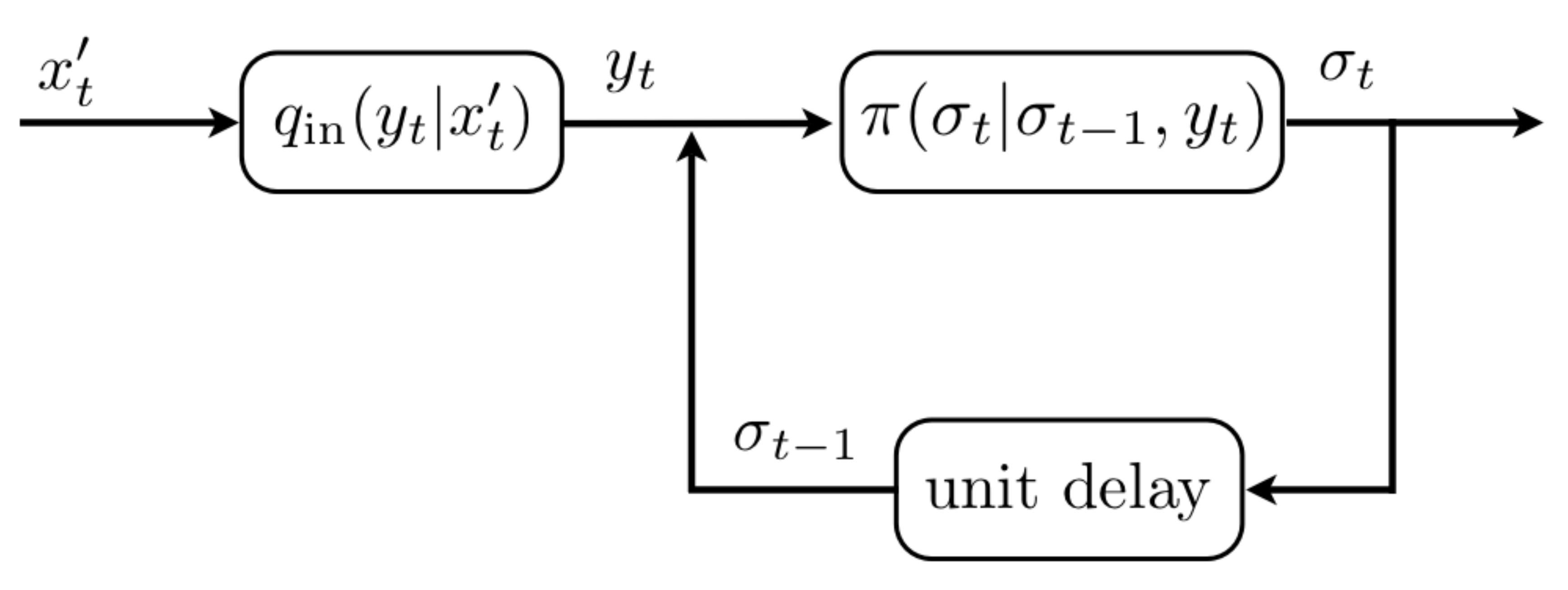}
\caption{\small An alternative representation to Fig.~\ref{fig:scheme}, where the strategy $\pi(\s_t|\s_{t-1},y_t)$ is viewed as a controller. The controller  receives both feedforward information $y_t$ through the sensor $q(y_t|x'_t)$, and feedback information $\s_{t-1}$, subject to delay (see also Table~\ref{fig:tablecorres}). \label{fig:control}}
\end{minipage} \hfill
\begin{minipage}[c]{.59\linewidth}
\centering
\centering
\includegraphics[width=\linewidth]{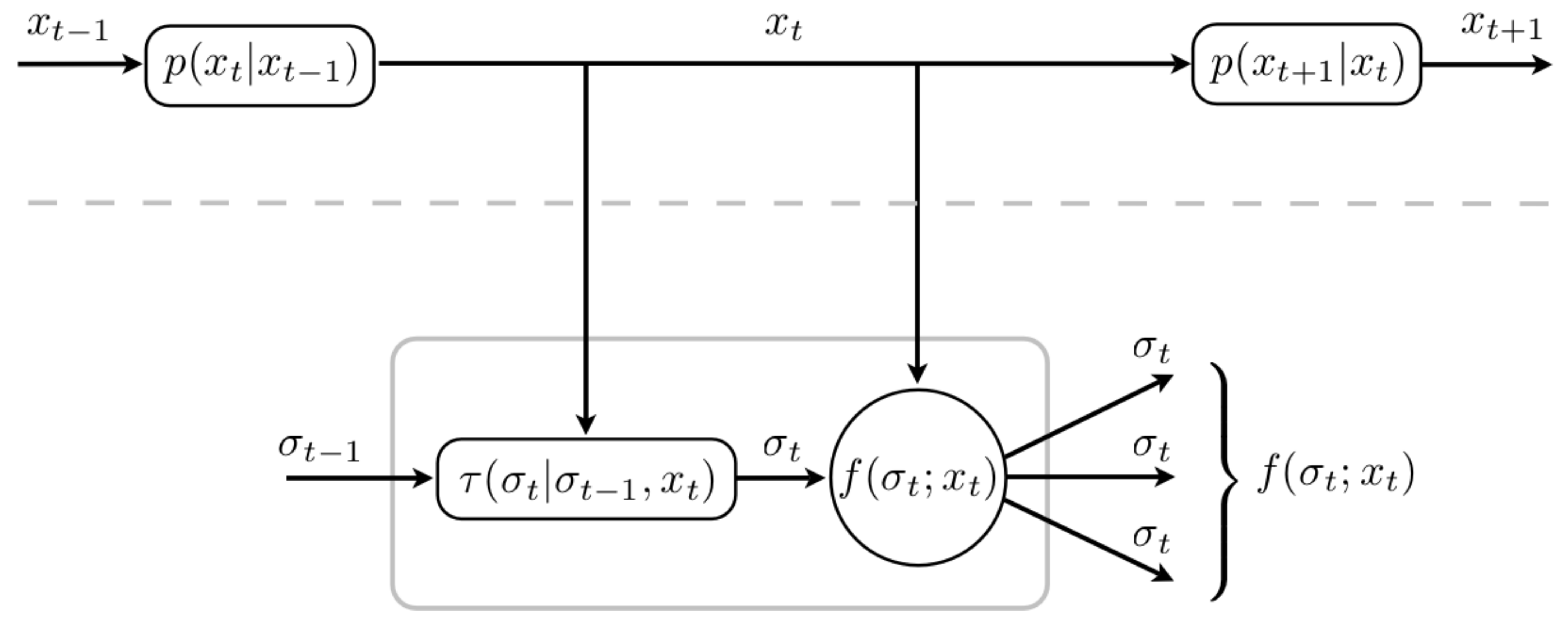}
\caption{\small Generalization of the model presented in Fig.~\ref{fig:scheme}, which corresponds to a
particular form of the transition matrix $\tau(\s_t|\s_{t-1},x_t)$. \label{fig:general}}
\end{minipage}
\end{figure}

Regulation is a general requirement for the sustainability and optimization of systems facing uncertainties. It forms the core issue of control in engineering, where acquired information is referred to as feedforward information and inherited information as feedback information (see Table~\ref{fig:tablecorres} and Fig.~\ref{fig:control}). Quantifying the value of limited information is a long-standing open conceptual problem in control theory~\cite{Witsenhausen71,Mitter01}. For growing populations, the law of large numbers and its extension, ergodicity, make the problem well-posed by introducing in the long-term, or infinite horizon limit in the language of control theory, an unambiguous loss-function, the growth rate of the population (see Sec.~\ref{sec:fitness}). Uncertainties are however generally not only due to limited information, and regulation must typically be made in presence of other constraints. These constraints can generally be classified in three categories: (i) constraints on estimation, i.e., on the acquisition of information about the current internal and external states, $\s_{t-1}$ and $x_t$, with the constraints on acquired information considered so far being an example; (ii) constraints on decision, i.e., on the computation of $\s_t$ from $\s_{t-1}$ and $y_t$; (iii) constraints on actuation, i.e., on the implementation of the switch from $\s_{t-1}$ to $\s_t$.  Biological constraints of the later type for instance arise when the types correspond to different developmental stages, in which case constraints of irreversibility are common~\footnote{Such constraints may be taken into account in our model by specifying a graph whose nodes are the types $\s_t$ and whose links are the possible transitions. An age structure can for instance be enforced by constraining the transitions matrices to have the form of Leslie matrices~\cite{Leslie45}. More generally, the constraints may restrict the graph of connectable types, as considered for instance in~\cite{KussellLeibler06}. Our model can also account for the presence of an unreliable "actuator" by constraining $\pi(\s_t|\s_{t-1},x_t)$ to be of the form $\pi(\s_t|\s_{t-1},y_t)=\sum_{\s_t'} a(\s_t|\s_t')\pi(\s_t'|\s_{t-1},y_t)$; this assumes that the inherited type $\s_t$ is an output of the actuator $a(\s_t|\s'_t)$, otherwise, if $\pi(\s_t|\s_{t-1},y_t)=\sum_{\s_t'} a(\s_t'|\s_t)\pi(\s_t|\s_{t-1},y_t)$ with $\s_t'$ controlling the multiplication rate $f(\s'_t;x_t)$ but not being inherited, the problem becomes equivalent to a model with effective multiplication rate $\tilde f(\s_t,x_t)=\sum_{\s'_t}f(\s'_t;x_t)a(\s'_t|\s_t)$.}. While constraints on the organisms limit the ability of a population to control its growth rate, it is interesting to notice that constraints on the environment, such as constraints on the possible states that may follow the current environmental state, render the future more predictable and have therefore the opposite effect of enhancing this ability~\footnote{Another intriguing duality, between control and knowledge, was noted by Shannon: "we may have knowledge of the past and cannot control it; we may control the future but have no knowledge of it".~\cite{Shannon59}}.\\

To encompass more general forms of constraints, our model can be extended to the model represented in Fig.~\ref{fig:general}, which considers a population with internal states $\s_t$ and environmental states $x_t$ described by
\beq
\N_t(\s_t)=f(\s_t;x_t)\sum_{\s_{t-1}}\tau(\s_t|\s_{t-1},x_t)\ \N_{t-1}(\s_{t-1}),
\eeq
where $\tau(\s_t|\s_{t-1},x_t)$ is a transition matrix, and where the environment follows as before a Markov chain $\pe(x_t|x_{t-1})$. Imposing constraints on control formally amounts to restricting $\tau(\s_t|\s_{t-1},x_t)$ to a subset $\C$ of the set of conceivable transition matrices. Different "information patterns"~\cite{Witsenhausen71}, specifying "who knows what and when", can thus be enforced. For instance, excluding feedback information corresponds to restricting $\tau(\s_t|\s_{t-1},x_t)$ to the form $\tau(\s_t|x_t)$, and excluding feedforward information to restricting it to the form $\tau(\s_t|\s_{t-1})$. The model with constraints on acquired information presented in Sec.~\ref{sec:model} can be formulated in this more general framework, by considering $\tilde x_t=(x_t,x'_t)$ for the environmental states, $\tilde \s_t=(\s_t,y_t)$ for the internal states, and $\tilde \pe(\tilde x_t|\tilde x_{t-1})=\qe(x'_t|x_t)\pe(x_t|x_{t-1})$ for the transition matrix between environmental states; the transition matrix $\tau(\tilde \s_t|\tilde \s_{t-1},\tilde x_t)$ must then be constrained to the form $\tau((\s_t,y_t)|(\s_{t-1},y_{t-1}),(x_t,x'_t))=\pi(\s_t|\s_{t-1},y_t)\qi(y_t|x'_t)$, which defines a subset $\C^{(\qi)}$ of admissible transition matrices. Several extensions of the model presented in Sec.~\ref{sec:model} can similarly be formulated. For instance, the sensor $\qi$ may be taken to depend on the type $\s_t$, thus allowing for different phenotypes to have different abilities to sense the environment: $\tau(\s_t|\s_{t-1},x_t)$ must then be constrained to the form $\tau(\s_t|\s_{t-1},x_t)=\sum_{y_t}\pi(\s_t|\s_{t-1},y_t)\qi(y_t|x_t,\s_{t-1})$. Another possible extension is to consider that the types are transmitted with some errors by constraining $\tau(\s_t|\s_{t-1};x_t)$ to the form $\tau(\s_t|\s_{t-1};x_t)=\sum_{z_t}\pi(\s_t|z_t,x_t)\mu(z_t|\s_{t-1})$, where $\mu(z_t|\s_{t-1})$ represents a given "mutational" transition matrix.\\

The questions (Q1) and (Q2) formulated in Sec.~\ref{sec:fitness} can be addressed in this more general framework by taking again the growth rate $\Lambda_{\pe;f}(\tau)$ as a fitness function. A first point of comparison is provided by $\hat\Lambda_{\pe;f}$, the optimal growth rate in absence of any constraint, obtained after optimization over $\tau$. The optimal transition matrix, $\hat \tau(\s_t|\s_{t-1};x_t)$, is easily characterized: at time $t$, it converts all the population to one of the types $\s$ maximizing $f(\s;x_t)$, irrespectively of the type $\s_{t-1}$ inherited from the previous generation. This optimal strategy is, however, generally excluded by the presence of constraints, characterized by the subset $\C$ to which $\tau(\s_t|\s_{t-1},x_t)$ must belong. Given $\C$, question (Q1) becomes the problem of finding a transition matrix $\hat \tau$ which maximizes $\Lambda_{\pe;f}(\tau)$ subject to the constraint $\tau\in\C$. This defines an optimal growth rate under constraints, $\hat \Lambda_{\pe;f}^{(\C)}$. The arguments of the previous sections can then be repeated {\it mutatis mutandis}. For intance, under the assumptions (A1) and (A3), the solution $\hat \tau$ is independent of $f$ and is associated with a generalization of Shannon's entropy, $H_{\pe}^{(\C)}\equiv\hat\Lambda_{\pe;f}-\hat \Lambda_{\pe;f}^{(\C)}$, given by
\beq\label{eq:H}
H_\pe^{(\C)}=\min_{\tau\in\C}\ \sum_x\pe(x)\ln \frac{1}{\tau(x;x)}.
\eeq
More generally, this quantity provides an upper bound for the cost of the constraints $\hat\Lambda_{\pe;f}-\hat \Lambda_{\pe;f}^{(\C)}$. On the other hand, question (Q2) pertains to the value of relaxing a constraint $\C$ to a lesser constraint $\C'\supset\C$, and amounts to estimating the quantity $I_{\pe;f}^{(\C;\C')}=\hat \Lambda_{\pe;f}^{(\C')}-\hat \Lambda_{\pe;f}^{(\C)}$, which generalizes the notion of mutual information $I_\pe^{(\qe,\qi)}$ obtained when $\C$ corresponds to the presence of the channels $(\qe,\qi)$, and $\C'$ to the absence of any channel.\\

The major problem not addressed in the present framework is the specification of the constraints and, more broadly, the characterization of the costs for implementing any particular strategy. For instance, when analyzing the value of acquired information, not only should we take into account the benefit provided by the communication channel $\qi$,  but also the cost for producing and operating it. This cost $c(\qi)$ is to be measured in terms of growth rate, and its value will determine whether the sensor $\qi$ has an adaptive value~\cite{KussellLeibler06}. More generally, a trade-off between cost and accuracy will arise if $c(\qi)$ is taken to be an increasing function of the accuracy of $\qi$. From this point of view, imposing constraints in the form of a subset $\C$ of achievable transition matrices corresponds to assuming that some strategies have infinite costs while some other are cost-less. Costs are also generally present not only in the estimation step, but also in the decision and actuations steps; for instance, there may be a cost for switching between types, as there are transaction costs in finance~\cite{IyengarCover00}.\\

Several other extensions can also be considered to explore other features of regulation in biological populations. For instance, from the standpoint of understanding the origin of diversification in a population, a key aspect of biological environments is their spatial heterogeneities. This feature may be incorporated at a mean-field level (not taking into account any geometrical properties of space) by making not only the acquired information $y_t$ specific to individuals, but also the environmental factor $z_t$ affecting their multiplication rates. We may thus assume that a "micro-environment" $(y_t,z_t)$ derives independently for each individual from the "macro-environment" $(x_t,x'_t)$, through a transition matrix $v(y_t,z_t|x_t,x'_t)$ attached to each individual. The dynamics of the population is then described by 
\beq\label{eq:heterog}
\N_t(\s_t)=\sum_{z_t}f(\s_t;z_t)\sum_{y_t,\s_{t-1}}\pi(\s_t|\s_{t-1},y_t)\ v(y_t,z_t|x_t,x'_t)\ \N_{t-1}(\s_{t-1}),
\eeq
where $x_t$ and $x_t'$ are again quenched environmental variables defined through $p(x_t|x_{t-1})$ and $\qe(x_t'|x_t)$.
We recover our previous model when $v(y_t,z_t|x_t,x'_t)=\qi(y_t|x'_t)\ \delta(z_t|x_t)$, i.e., $z_t=x_t$. More generally, if $y_t$ and $z_t$ are conditionally independent, i.e., $v(y_t,z_t|x_t,x'_t)=\qi(y_t|x_t)\ u(z_t|x_t)$ for some transition matrix $u(z_t|x_t)$, then the model can be reduced to a model without spatial heterogeneity but with an effective multiplication rate $\bar f(\s_t,x_t)=\sum_{z_t}f(\s_t,z_t)u(z_t|x_t)$. Note that this effective multiplication rate will generally be non-integer, even when $f(\s_t,z_t)$ represents an actual number of offsprings. $\bar f(\s_t;x_t)$ can also be non-diagonal even though $f(\s_t;x_t)$ is diagonal, so that in this case uncertainty is not measured by Shannon entropy even under the restrictive assumptions (A1), (A2), (A3). Note also that while the relevant temporal average of the multiplication rates is the geometric mean, the relevant spatial average in presence of spatial heterogeneities is an arithmetic mean. Another type of spatial heterogeneity is when several patches of population are present and each patch experiences independently an environmental sequence $\bar x$ described by the same Markov chain $\pe(x_t|x_{t-1})$. In the limit of infinitely many patches, the growth of the overall population is then not described by the quenched Lyapunov exponent $\Lambda_\text{quenched}=\lim_{t\to\infty}(1/t)\E\ln |\N_t|$ introduced in Sec.~\ref{sec:fitness}, but by the annealed Lyapunov exponent $\Lambda_\text{annealed}=\lim_{t\to\infty}(1/t)\ln\E|\N_t|$, which averages over all environmental sequences instead of focusing on typical ones. These two growth rates, sometimes called the "stochastic growth rate" and the "megamatrix growth rate" in the ecological literature~\cite{Tulja03}, satisfy the general relation $\Lambda_\text{quenched}\leq\Lambda_\text{annealed}$, which reflects the fact that with many independent patches, the overall population benefits from the few patches that experience atypical but particularly favorable environmental sequences.\\

Finally, we may mention briefly several other generalizations. A relatively straightforward one, which preserves a close connection to communication theory, is to consider continuous environmental and organismal states~\cite{HaccouIwasa95}; an interesting phenomenon of discretization whereby the optimal distribution of phenotype is actually discrete has then been described~\cite{SasakiEllner95}. The extension to continuous time is also relatively straightforward (see e.g.~\cite{KussellLeibler05}). Models where both time and space are continuous are also commonly considered in finance, and can be treated with the tools of stochastic calculus~\cite{Merton69}. Another kind of generalization is to introduce a longer time scale at which the transmission of the matrix $\pi$ is itself subject to mutations, thus allowing to address the issue of the evolution of $\pi$ towards $\hat \pi$. Finally, a more challenging extension is to account for interactions between individuals. For instance, it would be interesting to consider situations where the environment of one population is determined by another population, and to include the possibilities of communication between individuals, or of sexual reproduction. 

\section{Conclusion}

Applications of the concepts of information theory to biology have often been criticized on two main grounds~\cite{GodfreySmith07}: their failure to account for the directionality of information (the statistical problem of causality), and their failure to account for the value of information (the semantic problem of meaning). Following treatments of analogous problems in engineering and finance, we presented and analyzed a model in which these two features could be integrated. The analysis revealed another limitation of the usual concepts of information theory: their failure to account for the different levels at which information may be processed in a population, which led us to new generalizations of the entropy and mutual information.

\section*{Acknowledgements}

We thank O. Feinerman, G. Iyengar and E. Kussell for comments and discussions. OR was supported by a Simons Foundation fellowship from the Rockefeller University.

\newpage

\appendix

\section{Definition and properties of the model}\label{SI:mbpre}

In mathematical terms, our model belongs to the class of Athreya-Karlin models of multi-type branching process in random environments~\cite{AthreyaNey72,KarlinTaylor75}. Without seeking the highest level of rigor and generality, it can be described as follows. Let the environmental process be a discrete time, stationary and ergodic, stochastic process $\bar x=(x_1,\dots,x_t,\dots)$ with a finite set of states. Let $\mathcal{S}$ be a finite set of  admissible internal states $\s$ (types) of the individuals in a population. Let $\xi^{(t)}_{\s|\s';i}$ be the random variable giving the number of offsprings of type $\s$ that a particular individual $i$ of type $\s'$ generates at time $t$. The reproductive process, identical and independent for each individual $i$, is described by the joint distribution $\P_t\left(\xi^{(t)}_{\s|\s'}\}\right)$, which is conditional on the environmental state $x_t$, and therefore dependent on $t$. To complete the definition, we may consider starting at time $t=1$ with a single individual in the given type $\s_0\in\mathcal{S}$, but the asymptotic results, conditional on non-extinction, will not depend on this initial composition. The number $Z_t(\s)$ of individuals of type $\s$ at time $t$ is a random variable in terms of which the branching process can recursively be defined as
\beq\label{eq:branching}
Z_t(\s)=\sum_{\s'\in\mathcal{S}}\sum_{i=1}^{Z_{t-1}(\s')}\xi^{(t)}_{\s|\s';i},
\eeq
where different values of $i$ correspond to different realizations of the same random variable.\\

Our most basic model assumes that the reproductive process has the particular form
\beq
\P_t\left(\xi^{(t)}_{\s|\s'}=\xi\right)=R(\xi|\s;x_t)\ \pi(\s|\s';x_t),
\eeq
where $R(\xi|\s;x_t)$ is generally a transition matrix, with $\xi\in\NN$. Eq.~\eqref{eq:dyn} is obtained by taking the expectation in Eq.~\eqref{eq:branching} with respect to the random variables $\bar\xi=\{\xi^{(t)}_{\s|\s';i}\}_{t,\s,\s',i}$ for a given environmental sequence $\bar x=(x_1,\dots,x_t,\dots)$: 
\beq
\N_t(\s)=\E_{\bar\xi}[Z_t(\s)|\bar x].
\eeq
$\N_t(\s)$ depends on $R(\xi|\s;x_t)$ only through the multiplication rates defined by
\beq\label{eq:rf}
f(\s;x)=\E_\xi[\xi\ R(\xi|\s;x)|x]=\sum_{\xi=0}^\infty \xi\ R(\xi|\s;x).
\eeq
Other properties, not considered here, such as the probability of extinction, may depend on the fluctuations in the number of offsprings.\\

In the case of a constant environment, only two events can happen with positive probability~\cite{AthreyaNey72,KarlinTaylor75}: either the population goes extinct, i.e., $|Z_t|\equiv\sum_\s Z_t(\s)=0$ for some $t$, or it explodes, i.e., $Z_t\to\infty$ with $t\to\infty$. There are therefore two essential questions: (1) What is the probability of extinction? (2) What is the growth rate in the case of explosion? The answer to these questions is contained in the matrix $\A$ given by $\A_{\s\s'}=\E_{\xi_{\s|\s'}}[\xi_{\s|\s'}]=f(\s)\pi(\s|\s')$. Assuming that $\A$ is irreducible and aperiodic, it follows from the Perron-Frobenius theorem that $\A$  has an unique largest real eigenvalue $\lambda=\exp(\Lambda)$ with a corresponding eigenvector $v$ having strictly positive components, which can be normalized so that $\sum_\s v_\s=1$. Let $Q(\s)$ be the probability of extinction when initiating the population with a single individual of type $\s$; then the answer to (1) is:
\beq
Q(\s)<1,\ \forall \s\quad{\rm if\ and\ only\ if}\quad \Lambda>0.
\eeq
When $\Lambda>0$, the branching process is said to be supercritical, and we assume that our model is in this regime to prevent almost sure extinction. For such processes, the answer to (2) is given by Kesten-Stigum theorem~\cite{KurtzLyons97}:
\beq\label{eq:compopop}
e^{-t\Lambda}Z_t \to Wv\quad {\rm with\ probability\ 1},
\eeq
where $W$ is a scalar random variable with the following property:
\beq
\P(W>0)>0,\quad{\rm if\ and\ only\ if}\quad \E\left[\sum_{\s,\s'}\xi_{\s|\s'}\max(0,\ln \xi_{\s|\s'}) \right]<\infty.
\eeq
We assume that our model satisfies this $X\log X$ condition, i.e., $\E_{\xi_{\s|\s'}}[\xi_{\s|\s'}\max(0,\ln \xi_{\s|\s'})]=\sum_\xi  (\xi\ln \xi)R(\xi|\s)\pi(\s|\s')<\infty$. Thus, under the assumption of non-extinction, the distribution of the population is asymptotically described by the distribution of the first moments $\N_t(\s)=\E_\xi[Z_t(\s)|x]$, i.e., by the dominant eigenvector $v$ of $\A$~\cite{KurtzLyons97}:
\beq
\lim_{t\to\infty}\frac{Z_t(\s)}{\sum_{\s'}Z_t(\s')}=v_\s\quad\textrm{almost surely conditionally on non-extinction of the population.}
\eeq

In the case of varying environments, there is no longer necessarily convergence of the composition of the population as in Eq.~\eqref{eq:compopop}, but the Lyapunov exponent for the product of random matrices $\A^{(t)}\dots \A^{(1)}$ still corresponds to the typical growth rate of growing populations: $\Lambda=\lim_{t\to\infty}\frac{1}{t}\ln\left(\sum_\s Z_t(\s)\right)$ almost surely, as indicated in Eq.~\eqref{eq:tanny}. The stability condition required for this result to hold is that, with probability one~\cite{Tanny81},
\beq
\limsup_{t\to\infty}\frac{1}{t}\E\left[\ln\min_\s\sum_{\s'}(\A^{(t)}\A^{(t-1)}\dots \A^{(1)})_{\s\s'}\right]=\limsup_{t\to\infty}\frac{1}{t}\ln \| \A^{(t)}\A^{(t-1)}\dots \A^{(1)}\|,
\eeq
where $\| M\|$ represents a matrix norm, for instance $\| M\|=\sum_{\s,\s'}|M_{\s,\s'}|$.

\section{Analytically solvable models}\label{app:solvable}

Here, we present the analysis of our model in a few simple cases where a solution can be obtained analytically. Beyond horse race models, which are defined by the assumptions (A1), (A2), (A3) introduced in Sec.~\ref{sec:model}, a general class of solvable model is when (A3) is relaxed to allow for non-zero multiplication rate of the form $f(\s;x)=f(x)$ if $\s=x$, and $f(\s;x)=\tf(x)<f(x)$ otherwise (horse race models correspond to the case where $\tf(x)=0$). Under the assumptions (A1) and (A2) that the environment is i.i.d. and that $\qi=\delta$, the mathematical simplicity of these models stems from the fact that $\pi(x|x')$ contributes only to one term indexed by $x$ in the following sum:
\beq
\Lambda^{(\qe,\qi)}_{f;\pe}(\pi)=\sum_{x,x'}\qe(x'|x)\pe(x)\ \ln\left(\tf(x)+(f(x)-\tf(x))\pi(x|x')\right).
\eeq
The case where $\qi$ is a binary erasure channel as defined in Fig.~\ref{fig:channels} has also the same property. A subclass of this class of model is when the organisms and the environment have only two states, as in Fig.~\ref{fig:BEC}, \ref{fig:BSC}, \ref{fig:quant}, and \ref{fig:quant}. We present the details of the analysis of this two-state model below, always assuming that the environment is i.i.d..

\subsection{Two-state generic model with no information}\label{app:2state}

In absence of information,
\beq
\Lambda^{(\eta,\eta)}_{f;\pe}(\pi)=\pe(1)\ln(f(1;1)\pi(1)+f(2;1)\pi(2))+\pe(2)\ln(f(1;2)\pi(1)+f(2;2)\pi(2)).
\eeq
It is convenient to introduce the variables
\beq\label{eq:g12}
\g_1=\frac{f(1;1)-f(2;1)}{f(2;1)},\qquad \g_2=\frac{f(2;2)-f(1;2)}{f(1;2)}
\eeq
which, without loss in generality, can be assumed to be positive. Using $\pi(1)+\pi(2)=1$, the expression for the growth rate then becomes
\beq
\Lambda^{(\eta,\eta)}_{f;\pe}(\pi)=\hat\Lambda^{(\delta,\delta)}_\pe+\pe(1)\ln\frac{1+\g_1\pi(1)}{1+\g_1}+\pe(2)\ln\frac{1+\g_2\pi(2)}{1+\g_2}.
\eeq
Since $\pi(1)+\pi(2)=1$, the optimization involves only one independent variable, say $\pi(1)$, subject to the constraints $0\leq \pi(1)\leq 1$. As a function of $\pe(1)$, we thus obtain the following solution:
\beq
\hat\Lambda_{f;\pe}^{(\eta,\eta)}=
\begin{cases}
\hat\Lambda_{f;\pe}^{(\delta,\delta)}-\pe(1)\ln(1+\g_1) & {\rm if\ }\pe_1\leq\pe_c^{(1)},\\
\hat\Lambda_{f;\pe}^{(\delta,\delta)}-H_\pe^{(\eta,\eta)}+\pe(1)\ln\left(1+\frac{\g_1}{(1+\g_1)\g_2}\right)+\pe(2)\ln\left(1+\frac{\g_2}{(1+\g_2)\g_1}\right) & {\rm if\ }\pe_c^{(1)}\leq\pe(1)\leq \pe_c^{(2)},\\
\hat\Lambda_{f;\pe}^{(\delta,\delta)}-\pe(2)\ln(1+\g_2) & {\rm if\ }\pe(1)\geq \pe_c^{(2)},
\end{cases}
\eeq 
\beq
\textrm{with}\qquad \pe_c^{(1)}=\frac{\g_2}{\g_1+\g_2+\g_1\g_2},\qquad\pe_c^{(2)}=1-\frac{\g_1}{\g_1+\g_2+\g_1\g_2}.
\eeq
The first and third cases correspond respectively to $\hat \pi(1)=0$ and $\hat \pi(1)=1$, when not switching is optimal, and the intermediate case to $\hat \pi(1)=[\pe(1)(\g_1+\g_2+\g_1\g_2)-\g_2]/(\g_1\g_2)$. The location of the transitions between the different cases is represented in Fig.~\ref{fig:quant} for $\gamma_1=\gamma_2$, and the optimal growth rate $\hat\Lambda^{(\eta,\eta)}_\pe$ as a function of $\pe(1)$ in Fig.~\ref{fig:lambda} for a particular choice of the parameters.

\subsection{Two-state diagonal model with a binary erasure channel}\label{app:bec}

We assume here that assumption (A3) holds, i.e., $f$ is diagonal, but consider that assumption (A2) does not hold, and $\qi$ is the binary erasure channel $q_e$ defined in Fig.~\ref{fig:channels}. The optimization problem to be solved is $H_{\pe}^{(\delta,q_e)}=-\max_\pi\Upsilon_\pe^{(q_e)}(\pi)$ with
\beq
\Upsilon_\pe^{(q_e)}(\pi)=\pe(1)\ln(\pi(1|1)(1-\epsilon) + \pi(1|*)\epsilon)+\pe(2)\ln(\pi(2|2)(1-\epsilon)+\pi(2|*)\epsilon).
\eeq
Clearly $\hat \pi(1|1)=\hat \pi(2|2)=1$, so that we have a single independent variable over which to optimize, say $\pi\equiv\pi(1|*)=1-\pi(2|*)$. If we introduce $\gamma=\epsilon/(1-\epsilon)$ then
\beq
\Upsilon_\pe^{(q_e)}(\pi)=\ln(1-\epsilon)+\pe(1)\ln(1+\g \pi)+\pe(2)\ln(1+\g(1-\pi)).
\eeq
This is formally equivalent to the optimization performed above in absence of information (this formal equivalence extends beyond two-state models). As a function of the level of noise $\varepsilon$, the solution for the binary erasure channel is
\beq
H_\pe^{(\delta,q_e)}=
\begin{cases}
-\min(\pe(1),\pe(2))\ln(1-\varepsilon) & {\rm if\ }\varepsilon\leq\varepsilon_c(\pe)\\
H^{(\eta,\eta)}_\pe-\ln(2-\varepsilon) & {\rm if\ }\varepsilon\geq\varepsilon_c(\pe)
\end{cases}
,\qquad\textrm{with}\quad \epsilon_c(\pe)=\max\left(\frac{1-2\pe(1)}{1-\pe(1)},\frac{1-2\pe(2)}{1-\pe(2)}\right).
\eeq
If, for instance, we assume that $\pe(1)\leq\pe(2)$, then $\varepsilon_c(\pe)=(1-2\pe(1))/(1-\pe(1))$ and $\hat \pi=0$, $H_\pe^{(\delta,q_e)}=-\pe(1)\ln(1-\epsilon)$ for $\varepsilon<\varepsilon_c(\pe)$. An illustration is given in Fig.~\ref{fig:BEC} where we compare for $\pe(1)=0.1$ the individual information $I_\pe^{(\delta,q_e)}=H^{(\eta,\eta)}_\pe-H^{(\delta,q_e)}_\pe$ with the mutual information $I_\pe^{(q_e,\delta)}=H^{(\eta,\eta)}_\pe-H_\pe^{(q_e,\delta)}=(1-\varepsilon)H^{(\eta,\eta)}_\pe$ for the same binary erasure channel.

\subsection{Two-state model diagonal with a binary symmetric channel}\label{app:bsc}

We assume here that assumption (A3) holds, i.e., $f$ is diagonal, but consider that assumption (A2) does not hold, and $\qi$ is the binary symmetric channel $q_s$ defined in Fig.~\ref{fig:channels}. The binary symmetric channel corresponds to $q_s(1|1)=q_s(2|2)=1-\varepsilon$ and $q_s(2|1)=q_s(2|1)=\varepsilon$, where, without loss in generality, we can assume that $0\leq\varepsilon\leq 1/2$. The optimization problem to be solved is $H^{(\delta,q_s)}_{\pe}=-\max_\pi\Upsilon_\pe^{(q_s)}(\pi)$ with
\beq
\Upsilon_\pe^{(q_s)}(\pi)\equiv\pe(1)\ln(\pi(1|1)(1-\varepsilon)+\pi(1|2)\varepsilon)+\pe(2)\ln(\pi(2|2)(1-\varepsilon)+\pi(2|1)\epsilon).
\eeq
We have here two independent parameters over which to optimize, $\pi_1=\pi(1|1)$ and $\pi_2=\pi(2|2)$, since $\pi(2|1)=1-\pi(1|1)=1-\pi_1$ and  $\pi(1|2)=1-\pi(2|2)=1-\pi_2$. If we introduce $\gamma=\varepsilon/(1-\varepsilon)$, the function to optimize becomes
\beq
\Upsilon_\pe^{(q_s)}(\pi_1,\pi_2)\equiv\ln(1-\varepsilon)+\pe(1)\ln(\pi_1+(1-\pi_2)\g)+\pe(2)\ln(\pi_2+(1-\pi_1)\g).
\eeq
The calculation shows that the only case where we can have both $0<\hat \pi_1<1$ and $0<\hat \pi_2<1$ is the blind case where $\g=1$ and $\varepsilon=1/2$, for which we have the proportional betting solution $\hat \pi_1=\pe(1)$ and $\hat \pi_2=\pe(2)$. In any other case, $\hat \pi_1=1$ or $\hat \pi_2=1$, which reduces the problem to an optimization over a single variable. The solution is
\beq
H_\pe^{(\delta,q_s}=
\begin{cases}
-\ln(1-\varepsilon) & {\rm if\ }\varepsilon\leq\varepsilon_c(\pe)\\
H^{(0)}_\pe-\varepsilon_c(\pe)\ln((1-\varepsilon)/\varepsilon) & {\rm if\ }\varepsilon\geq\varepsilon_c(\pe)
\end{cases}
,\qquad\textrm{with}\quad \varepsilon_c(\pe)=\min(\pe(1),\pe(2)).
\eeq
If for instance $\pe(1)\leq\pe(2)$, when $\varepsilon_c(\pe)<\pe(1)$ we have both $\hat \pi_1=1$ and $\hat \pi_2=1$ while for $\varepsilon_c(\pe)>\pe(1)$, we have $\hat \pi_2=1$ but $\hat \pi_1<1$. An illustration is given in Fig.~\ref{fig:BSC} where we compare for $\pe(1)=0.1$ the individual information $I_\pe^{(\delta,q_s)}$ with the mutual information $I_\pe^{(q_s,\delta)}$ for the same binary symmetric channel. 

\section{A solvable model in non i.i.d.~environments}\label{sec:timing}

We connect here the model proposed in~\cite{KussellLeibler05} to the framework of this paper and discuss how uncertainties in timing can thus be quantified. We will thus make explicit the time scales involved in the trade-off between short-term adjustment to the current environmental conditions, and longer term anticipation of changes of these conditions.\\

Two time scales govern short-term adjustement: the time $\tau(\e)$ that an environmental state $\e$ lasts, and the time $\a(\e;\e')$ that it takes for the type with largest multiplication rate in $\e$ to dominate the population; this later adjustment time depends on the composition of the population at the time of the environmental change, and the notation $\a(\e;\e')$ indicates that we consider a population initially adjusted to some other environmental state $\e'$. In the so-called adiabatic regime where the population has always time to adjust to the current environment, i.e., $\a(\e;\e')\ll\tau(\e)$ for all $\e$ and $\e'\neq\e$, the dynamics of the population has a common feature with horse race models: at the end of an environmental period, most of the population shares a common type, much as in horse race models where, at the end of a time step, only the money invested in the winning horse yields a non-zero payoff. In horse race models, the particular form of the multiplication rates, where $f(\s;\e)=0$ whenever $\s\neq\e$, implies that the uncertainty cost $\hat\Lambda_{\w;f}-\hat\Lambda_{\w;f}^{(\C)}$ can be measured by the entropic function $\hat H_\w^{(\C)}$ defined in Eq.~\eqref{eq:H}, where the transition matrix for the environment is here denoted $\omega(\e_t|\e_{t-1})$. We shall see that the same function contributes to the uncertainty cost of systems in the adiabatic regime; in particular, the entropy of the environmental process can account for part of the uncertainty cost, as first noticed in~\cite{KussellLeibler05}.\\

The characteristic time that an environmental state $\epsilon$ lasts can be defined as the mean time $\tau(\e)$ spent in $\e$
\beq
\tau(\e)=\frac{1}{1-\omega(\e|\e)}.
\eeq
To simply define an adjustment time $\a(\e;\e')$, we assume that each environmental state $\e$ is associated with a different optimal type denoted with the same symbol $\s=\e$, i.e., $f(\e;\e)> f(\s;\e)$ for all $\s\neq\e$. We also assume that, in the course of a single time step, an individual is more likely to stay in its current type than to adopt a new one, i.e., $\pi(\s'|\s;\e)\ll 1$ for $\s'\neq\s$; these two assumptions ensure that, in a constant environment $\e$, an optimal type $\s=\e$ can indeed dominate the population if given sufficient time. Under these assumptions, a population initially composed of $\N_0$ individuals adjusted to environment $\e'$, has, after a time $t$ spent in environment $\e$, a number $\N_t(\e)$ of individuals of type $\e$ which is given by
\beq\label{eq:approx}
\N_t(\e)\simeq f(\e;\e)^tQ(\e;\e')\N_0.
\eeq 
Here, $Q(\e;\e')$, which satisfies $0<Q(\e;\e')<1$, can be interpreted as a (non-symmetric) overlap between the compositions of the population before and after the environmental change from $\e'$ to $\e$; as shown in appendix~\ref{SI:adiab}, it is given for $\e\neq\e'$ by
\beq\label{eq:q}
Q(\e;\e')=\Delta(\e;\e')\pi(\e|\e';\e')+\Delta(\e';\e)\pi(\e|\e';\e)\qquad\textrm{with}\quad\Delta(\e;\e')=\frac{f(\e;\e')}{f(\e';\e')-f(\e;\e')}.
\eeq
The adjustment time $\alpha(\e;\e')$ can then be defined as the time at which the sub-population of type $\e$ starts to overtake the sub-population of type $\e'$, $\N_t(\e)\sim \N_t(\e')$; given Eq.~\eqref{eq:approx} and $\N_t(\e')\sim f(\e';\e)^t\N_0$, this leads to 
\beq
\alpha(\e;\e')=\frac{1}{\ln f(\e;\e)-\ln f(\e;\e')}\ln\left(\frac{1}{Q(\e;\e')}\right).
\eeq

In the "adiabatic regime" where environmental periods exceed the adjustment times, i.e., $\alpha(\e;\e')\ll\tau(\e)$ for all $\e$ and $\e'\neq\e$, we obtain from Eq.~\eqref{eq:approx} a simple expression for the Lyapunov exponent (see appendix~\ref{SI:adiab}),
\beq\label{eq:Ladiab}
\Lambda_{\w;f}^\textrm{(adiabatic)}(\pi)\simeq\sum_\e\w(\e)\ln f(\e;\e)-\sum_{\e,\e'}\w(\e|\e')\w(\e')\ln\left(\frac{1}{Q(\e;\e')}\right),
\eeq
with the convention that, when $\e'\neq\e$, $Q(\e;\e)=\pi(\e|\e;\e)$.
The first term on the right hand side corresponds to the optimal Lyapunov exponent, $\hat\Lambda_{\w;f}$, and the second, when optimized over $\pi$, to the uncertainty cost $\hat \Lambda_{\w;f}-\hat\Lambda_{\w;f}^{(\C)}$. This second term depends, via $Q(\e;\e')$, on both the transition matrix $\pi$, and the values of the multiplication rates $f$. These two contributions are, however, set apart when the transition matrix $\pi$ can be factorized in Eq.~\eqref{eq:q}, which occurs in two notable cases. One case is in absence of a sensor, when $\pi(\e|\e';\e')=\pi(\e|\e';\e)=\pi(\e|\e')$, and, therefore,
\beq
Q(\e;\e')=\Gamma(\e;\e')\pi(\e|\e')\qquad\textrm{with}\quad\Gamma(\e;\e')=\Delta(\e;\e')+\Delta(\e';\e).
\eeq
Another case is in presence of a reliable sensor, when $\pi(\e|\e';\e')\ll \pi(\e|\e';\e)$, and, therefore,
\beq
Q(\e;\e')\simeq\Gamma(\e;\e')\pi(\e|\e';\e)\qquad\textrm{with}\quad\Gamma(\e;\e')=\Delta(\e';\e).
\eeq
In both of these cases, the second term of the right-hand side of Eq.~\eqref{eq:Ladiab} can be decomposed as
\beq\label{eq:withlastterm}
\sum_{\e,\e'}\w(\e|\e')\w(\e')\ln\left(\frac{1}{Q(\e;\e')}\right)=\sum_{\e,\e'\neq\e}\w(\e|\e')\w(\e')\ln\Gamma(\e;\e')-\sum_{\e,\e'}\w(\e|\e')\w(\e')\ln \pi(\e|\e';\e).
\eeq
The term $-\sum_{\e,\e'}\w(\e|\e')\w(\e')\ln \pi(\e|\e';\e)$ is analogous to the term obtained for a 
horse-race model in a Markov environment. The other term, involving $\Gamma(\e;\e')$, may be interpreted as the cost of the delay for transferring the majority of the population from one type to the next (such a term is absent in horse race models where transfers of capital can occur instantaneously prior to the environmental change).\\

Since only the term $-\sum_{\e,\e'}\w(\e|\e')\w(\e')\ln \pi(\e|\e';\e)$ depends on the transition matrix $\pi$, the optimal strategy is the one that minimizes it, and it has exactly the same features as in horse race models. If we consider for instance the situation with no information, we find $\hat \pi^{(\eta)}(\e|\e')=\w(\e|\e')$, the proportional betting strategy, and the optimal value of the last term in Eq.~\eqref{eq:withlastterm} is the entropy rate
\beq
H_\w^{(\eta)}=-\sum_{\e,\e'}\w(\e|\e')\w(\e')\ln\w(\e|\e').
\eeq
It is instructive to make here explicit the characteristic times $\tau(\e)$ giving the mean duration in each environmental state $\e$. This is done by introducing $\tilde\w(\e|\e')=\w(\e|\e')/\tau(\e')$, the probability that the environment changes from state $\e'$ to state $\e\neq\e'$, given that it does change its state, and $\tilde \w(\e)=\w(\e)\tau(\e)/\tau$, the probability to end up in state $\e$ when such an environmental change occurs, with $\tau=\sum_\e \tilde \w(\e)\tau(\e)$ representing the mean duration of a period of constant environment. With these definitions, it can indeed be shown that
\beq\label{eq:decompH}
H^{(\eta)}_{\w}=\frac{1}{\tau} H^{(\eta)}_{\tilde \w}+\sum_{\e}\frac{\tau(\e)}{\tau}\tilde \w(\e) H^{(\eta)}_{b(1/\tau(\e))}
\eeq
where $b(1/\tau(\e))$ refers to the Bernoulli distribution with parameter $1/\tau(\e)$ whose entropy is
\beq
 H^{(\eta)}_{b(1/\tau(\e))}=-\frac{1}{\tau(\e)}\ln\frac{1}{\tau(\e)}-\left(1-\frac{1}{\tau(\e)}\right)\ln\left(1-\frac{1}{\tau(\e)}\right).
\eeq
Eq.~\eqref{eq:decompH} shows that the uncertainty has two components, each of which measurable by an entropy: an uncertainty about the nature of the next environment, captured by $H^{(\eta)}_{\tilde\w}$, and an uncertainty about the timing of environmental changes, captured by $H^{(\eta)}_{b(1/\tau(\e))}$. As in horse race models, the maladjustment cost has, in the "adiabatic" limit and in absence of information, the form of a relative entropy which can also be decomposed in two terms; when the corresponding expressions are expanded for large $\tau(\e)$, the formulas presented in~\cite{KussellLeibler05} are thus recovered.

\section{Proof of the entropic bound}\label{SI:Hbound}

We prove here the bound on the uncertainty cost,
\beq
\hat\Lambda_{\pe;f}^{(\delta,\delta)}-\hat\Lambda_{\w;f}^{(\qe,\qi)}\leq H_\pe^{(\qe,\qi)},
\eeq
for i.i.d.~environments but arbitrary multiplication rates $f$.\\

By defining $\e=(x,x')$, $\w(\e)=\qe(x'|x)\pe(x)$ and $q(y|\e)=\qi(y|x')$, the Lyapunov exponent $\Lambda_{\pe;f}^{(\qe,\qi)}(\pi)$ is more concisely, but equivalently, written
\beq\label{eq:Lconcise}
\Lambda_{\omega;f}^{(q)}(\pi)=\sum_\e\omega(\e)\ln\left(\sum_{\s,y} f(\s;\e)\pi(\s|y)q(y|\e)\right).
\eeq
For each environmental state $\e$, let $\phi(\e)$ be one of the types $\s$ with maximal multiplication rate, such that $f(\s;\e)\leq f(\phi(\e);\e)$ for all $\s$. An assumption is here that $f(\phi(\e);\e)>0$ for all $\e$, which is necessary for the population not to go extinct, assuming that $\w(\e)>0$ for all $\e$.\\

From the definition of $\hat\Lambda_{\w;f}^{(q)}$ as $\max_\pi \Lambda_{\w;f}^{(q)}(\pi)$, for any probability matrix $\pi(\s|y)$, we have
\begin{align}
\hat\Lambda_{\w;f}^{(\delta)}-\hat\Lambda_{\w;f}^{(q)}
&\leq\sum_\e\w(\e)\ln f(\phi(\e);\e)-\sum_\e\w(\e)\ln\left(\sum_{\s,y} f(\s;\e)\pi(\s|y)q(y|\e)\right)\\
&=-\sum_\e \w(\e)\ln\left(\sum_{\s,y} \frac{f(\s;\e)}{f(\phi(\e);\e)}\pi(\s|y)q(y|\e)\right)\\
&=-\sum_\e\w(\e)\ln\left(\sum_y\left(\pi(\phi(\e)|y)+\sum_{\s\neq\phi(\e)}\frac{f(\s;\e)}{f(\phi(\e);\e)}\pi(\s|y)\right)q(y|\e)\right)\\
&\leq-\sum_\e\w(\e)\ln\left(\sum_y \pi(\phi(\e)|y)q(y|\e)\right)\label{eq:last}.
\end{align}
Let $\hat \rho(\e|y)$ be a transition matrix that achieves the minimum in the definition of $H_\w^{(q)}$, which is
\beq\label{eq:Hwq}
H_\w^{(q)}=\min_\rho\sum_\e\w(\e)\ln\left(\sum_y \rho(\e|y)q(y|\e)\right)^{-1}.
\eeq
Under the assumption that $\phi$ is injective, by taking $\pi(\s|y)=\hat \rho(\e|y)$ if $\s=\phi(\e)$, and $\pi(\s|y)=0$ if there is no $\e$ for which $\s=\phi(\e)$, we define a probability matrix $\pi$ for which the right-hand side of Eq.~\eqref{eq:last} corresponds exactly to $H_\w^{(q)}$. Hence
\beq
\hat\Lambda_{\w;f}^{(\delta)}-\hat\Lambda_{\w;f}^{(q)}\leq H_\w^{(q)}.
\eeq

If $\phi$ is non-injective, an even tighter upper bound can be designed. To this end, we go back to the variables $(x,y')$ and define a coarse-grained environmental process whose states are the equivalent classes for the relation $\phi(x)=\phi(z)$, and whose probability distribution is defined on the quotient set by $\tilde \pe(\tilde x)=\sum_{z\in\tilde x}\pe(z)$ for every equivalent class $\tilde x=\{z:\phi(z)=\phi(x)\}$. Introducing also $\tilde q_{\rm env}(x'|\tilde x)=\sum_{x\in\tilde x}\qe(x'|x)\pe(x)/\tilde\pe(\tilde x)$, the expression in Eq.~\eqref{eq:last} becomes
\beq
\sum_{x',x}\qe(x'|x)\pe(x)\left(\sum_y \pi(\phi(x)|y)q(y|x')\right)=\sum_{x',\tilde x}\tilde q_{\rm env}(x'|\tilde x)\tilde\pe(\tilde x)\ln\left(\sum_y \pi(\phi(\tilde x)|y)q(y|x')\right),
\eeq
where $\tilde \pi(\phi(\tilde x)|y)=\pi(\phi(x)|y)$ is well-defined by definition of the equivalence relation. We are then reduced to the injective case, and can therefore conclude
\beq
\hat\Lambda_{\pe;f}^{(\delta,\delta)}-\hat\Lambda_{\pe;f}^{(\qe,q)}\leq H_{\tilde \pe}^{(\tilde \qe,q)}.
\eeq
Finally, it follows from the definition of generalized entropy $H_{\pe}^{(\qe,q)}$ and from the concavity of the logarithm that coarse-graining always reduces the entropy, i.e., $H_{\tilde \pe}^{(\tilde \qe,q)}\leq H_{\pe}^{(\qe,q)}$, thus proving the entropic bound in the general case.

\section{Proof of the maladjustment bound}\label{SI:Dbound}

We prove here a bound on the cost incurred for following a non-optimal strategy. This bound generalizes the bound established for models of financial investments where $\qi=\delta$~\cite{CoverThomas91}. We consider here an i.i.d.~environment but arbitrary multiplication rates $f$. If $\hat \pi_{\pe'}$ denotes an optimal strategy for the i.i.d.~environment with probability $\pe'(x)$ rather than $\pe(x)$, we show that 
\beq\label{eq:Dbound}
\hat\Lambda_{\pe;f}^{(\qe,\qi)}-\Lambda_{\pe;f}^{(\qe,\qi)}(\hat \pi_{\pe'})\leq D(\pe\|\pe'),
\eeq
where $D(\pe\|\pe')=\sum_x\pe(x)\ln[\pe(x)/\pe'(x)]$ is the relative entropy between the environmental distributions $\pe$ and $\pe'$.\\

Writing the Lyapunov exponent $\Lambda_{\pe;f}^{(\qe,\qi)}(\pi)$ as in Eq.~\eqref{eq:Lconcise}, we show that
\beq
\hat\Lambda_{\omega;f}^{(q)}-\Lambda_{\omega;f}^{(q)}(\hat \pi_{\omega'})\leq D(\omega\|\omega'),
\eeq
from which Eq.~\eqref{eq:Dbound} follows by taking $\e=(x,x')$ and $\omega(\e)=\qe(x'|x)\pe(x)$ and $\omega'(\e)=\qe(x'|x)\pe'(x)$.\\

The proof relies on the characterization of $\hat \pi_{\omega'}$ in terms of the so-called Kuhn-Tucker conditions~\cite{CoverThomas91}, which generalizes to inequality constraints the method of Lagrange multipliers. These conditions imply here the existence of a set of $\lambda_y\geq 0$ satisfying
\beq\label{eq:KT}
\sum_\e\omega'(\e)\left(\frac{f(\s;\e)q(y|\e)}{\sum_{y',\s'}f(\s';\e)\hat \pi_{\omega'}(\s'|y')q(y'|\e)}\right)
\begin{cases}
= \lambda_y & {\rm if\ }\hat \pi_{\omega'}(\s|y)>0,\\
\leq\lambda_y & {\rm if\ } \hat \pi_{\omega'}(\s|y)=0,
\end{cases}
\eeq
and $\sum_y\lambda_y=1$.\\

After noticing that by taking the union of the two environmental state spaces if necessary, we can assume that the two processes described by $\omega$ and $\omega'$ have same states, we generalize a proof presented in~\cite{CoverThomas91} by considering the following series of inequalities:
\beq
\begin{split}
\hat\Lambda^{(q)}_{\omega;f}-\Lambda^{(q)}_{\omega;f}(\hat \pi_{\omega'})&=\sum_\e\omega(\e)\ln\left(\frac{\sum_{y,\s}f(\s;\e)\hat \pi_\omega(\s|y)q(y|\e)}{\sum_{y',\s'}f(\s';\e)\hat \pi_{\omega'}(\s'|y')q(y'|\e)}\right)\\
&=\sum_\e\omega(\e)\ln\left(\frac{\omega'(\e)\sum_{y,\s}f(\s;\e)\hat \pi_\omega(\s|y)q(y|\e)}{\omega(\e)\sum_{y',\s'}f(\s';\e)\hat \pi_{\omega'}(\s'|y')q(y'|\e)}\right)+D(\omega\|\omega')\\
&\leq \ln\left(\sum_\e\frac{\omega'(\e)\sum_{y,\s}f(\s;\e)\hat \pi_\omega(\s|y)q(y|\e)}{\sum_{y',\s'}f(\s';\e)\hat \pi_{\omega'}(\s'|y')q(y'|\e)}\right)+D(\omega\|\omega')\\
&= \ln\left(\sum_{y,\s}\hat \pi_\omega(\s|y)\sum_\e\omega'(\e)\left(\frac{f(\s;\e)q(y|\e)}{\sum_{y',\s'}f(\s';\e)\hat \pi_{\omega'}(\s'|y')q(y'|\e)}\right)\right)+D(\omega\|\omega')\\
&\leq  \ln\left(\sum_{y,\s}\hat \pi_\omega(\s|y)\lambda_y\right)+D(\omega\|\omega')\\
&=  \ln\left(\sum_y\lambda_y\right)+D(\omega\|\omega')\\
&=D(\omega\|\omega'),
\end{split}
\eeq
where the first inequality follows from the concavity of the logarithm (Jensen's inequality), and the second from Eq.~\eqref{eq:KT}.

\section{Perturbative approximation}\label{SI:adiab}

With $\N_t$ representing the population vector whose components $\N_t(\s)$ are the mean number of individuals of type $\s$, and assuming here that no information is acquired, Eq.~\eqref{eq:A} can be written
\beq
\N_{t+1}=\left(\A^{(\e_t)}_0+\A^{(\e_t)}_1\right)\N_t
\eeq
where, using a braket notation, the elements of the matrices $\A^{(\e)}_0$ and $\A^{(\e)}_1$, are
\begin{align}
&\langle \s'|\A^{(\e)}_0|\s\rangle=
\begin{cases}
f(\s;\e) & {\rm if\ }\s'=\s,\\
0 & {\rm if\ }\s'\neq \s,
\end{cases}\\
&\langle \s'|\A^{(\e)}_1|\s\rangle=
\begin{cases}
-f(\s;\e)(1-\pi(\s|\s;\e))& {\rm if\ }\s'=\s,\\
f(\s';\e) \pi(\s'|\s;\e)& {\rm if\ }\s'\neq \s.
\end{cases}
\end{align}
The rational for this decomposition is that $\A_1$ is a perturbation for $\A_0$ when the assumption is made that individuals are less likely to switch to a new type than to retain their current type. To simplify the discussion, we also assume that each environmental state $\e$ is associated with an unique optimal type $\s=\e$, satisfying $f(\e;\e)>f(\s;\e)$ for all $\s\neq\e$.\\

Under these assumptions, we can derive approximate expressions for the eigenvalues and eigenvectors of $\A_1+\A_0$ by a perturbative expansion. If $\lambda_\s^{(\e)}$, $|\psi_\s^{(\e)}\rangle$ and $\langle\psi_\s^{(\e)}|$ denote respectively the eigenvalue, and the right and left eigenvectors of the matrix $\A^{(\e)}_0+\A^{(\e)}_1$, we have, to first order in the perturbative expansion:
\begin{align}
\lambda_\s^{(\e)}&=f(\s;\e)\pi(\s|\s;\e),\\
|\psi_\s^{(\e)}\rangle&=|\s\rangle+\sum_{\s'\neq \s}\frac{f(\s';\e)}{f(\s;\e)-f(\s';\e)}\pi(\s'|\s;\e)|\s'\rangle,\\ 
\langle\psi_\s^{(\e)}|&=\langle \s|+\sum_{\s'\neq \s}\frac{f(\s';\e)}{f(\s;\e)-f(\s';\e)}\pi(\s|\s';\e)\langle \s'|.
\end{align}
The domain of validity of this approximation can be estimating by comparing the first and second order contributions to the eigenvalues, and we thus get the condition
\beq\label{eq:condition}
\left|\sum_{\s'\neq \s}\frac{f(\s;\e)f(\s';\e)}{f(\s;\e)-f(\s';\e)}\pi(\s'|\s;\e)\pi(\s|\s';\e)\right|\ll f(\s;\e)(1-\pi(\s|\s;\e)),\quad\forall \s,\e.
\eeq 
Given that $1-\pi(\s|\s;\e)=\sum_{\s'\neq\s}\pi(\s'|\s;\e)$, a sufficient condition for Eq.~\eqref{eq:condition} to hold is
\beq
\pi(\s|\s';\e)\ll \left|\frac{f(\s;\e)-f(\s';\e)}{f(\s';\e)}\right|,\quad\forall \s,\s',\e\quad (\s\neq \s').
\eeq
This shows that the underlying assumption behind the perturbative expansion is that changes in composition of the population should primarily be due to differences in multiplication rates, rather than be due to switches to new types.\\

For the dynamics to be in the "adiabatic regime", it is furthermore necessary that the environment stays long enough in any given state $\e$. When this is the case, the population vector $\N_t$ is, at the end of a period spent in state $\e$, quasi aligned along the dominant eigenvector vector  $|\psi_\e^{(\e)}\rangle$ that corresponds to the most favorable type for environment $\e$, $|\psi_\e^{(\e)}\rangle\simeq |\e\rangle$. If $\e'$ was the environmental state preceding the current state $\e$, the system is described by $|\psi_{\e'}^{(\e')}\rangle$ at $t=0$ and, at $t=\tau(\e)$, when the environmental state becomes $\e$, we require that
\beq\label{eq:adiab}
\left|\sum_{\s\neq \e}\langle\psi_\s^{(\e)}|\psi_{\e'}^{(\e')}\rangle\left(\lambda_\s^{(\e)}\right)^{\tau(\e)}\right|
\ll \langle\psi_\e^{(\e)}|\psi_{\e'}^{(\e')}\rangle\left(\lambda_\e^{(\e)}\right)^{\tau(\e)}.
\eeq
This condition can be made explicit by using the perturbative formulas
\beq
\langle\psi_\s^{(\e)}|\psi_{\e'}^{(\e')}\rangle=
\begin{cases}
1 & {\rm if\ }\s=\e',\\
\frac{f(\s;\e')}{f(\e';\e')-f(\s;\e')}\pi(\s|\e';\e')+\frac{f(\e';\e)}{f(\s;\e)-f(\e';\e)}\pi(\s|\e';\e) & {\rm if\ }\s\neq \e'.
\end{cases}
\eeq
Given $\e$, the longest delay time $\alpha(\e;\e')$ is therefore when the preceding environment $\e'$ corresponds to the second largest eigenvalue of $\A^{(\e)}_1$, that is, when $\e'=\s$ such that $f(\s;\e)=\max_{\s'\neq \e}f(\s';\e)$. Denoting $Q(\e;\e')=\langle\psi_\e^{(\e)}|\psi_{\e'}^{(\e')}\rangle$ we thus obtain the condition $f(\e';\e)^{\tau(\e)}\ll f(\e;\e)Q(\e;\e')$ or, equivalently,
\beq
\tau(\e)\ll\frac{1}{\ln f(\e;\e)-\ln f(\e';\e)}\ln\frac{1}{Q(\e;\e')},
\eeq
where the right-hand side can be taken as a definition for the adjustment time $\alpha(\e;\e')$.\\

Let now denote $\tilde\w(\e|\e')$ the probability for the environment to change from state $\e'$ to state $\e\neq\e'$, {\it given that it does change its state}. This is given by $\tilde\w(\e|\e')=\w(\e|\e')/(1-\w(\e'|\e'))$, where $\tau(\e')=1/(1-\w(\e'|\e'))$ also corresponds to the mean time spent in state $\e'$. The unconditional probability to end up in state $\e$ {\it when an environmental change occurs} is $\tilde\w(\e)=\w(\e)\tau/\tau(\e)$ where $\tau=\sum_\e \tilde w(\e)\tau(\e)$ represents the mean duration of a period of constant environment. In terms of these quantities, the growth rate is
\beq
\Lambda^{(\rm adiabatic)}_\w(\pi)=\frac{1}{\tau}\sum_{\e,\e'\neq\e}\tilde w(\e|\e')\tilde w(\e')\ln \left(\left(\lambda_\e^{(\e)}\right)^{\tau(\e)}\langle\psi_\e^{(\e)}|\psi_{\e'}^{(\e')}\rangle\right),
\eeq
which is also equivalent to
\beq
\Lambda^{(\rm adiabatic)}_\w(\pi)=\sum_{\e,\e'}\w(\e|\e')w(\e')\ln f(\e;\e)-\sum_{\e,\e'}\w(\e|\e')w(\e')\ln \frac{1}{Q(\e;\e')},
\eeq
with the convention that $Q(\e;\e)=\pi(\e|\e;\e)$, while $Q(\e;\e')$ for $\e\neq\e'$ is given by Eq.~\eqref{eq:q}.


\begin{thebibliography}{10}

\bibitem{MaynardSmith00}
J.~Maynard-Smith.
\newblock The concept of information in biology.
\newblock {\em Philosophy of Science}, 67(2):177--194, 2000.

\bibitem{Jablonka02}
E.~Jablonka.
\newblock Information: Its interpretation, its inheritance, and its sharing.
\newblock {\em Philosophy of Science}, 69(4):578--605, 2002.

\bibitem{Nurse08}
P.~Nurse.
\newblock Life, logic and information.
\newblock {\em Nature}, 454:424--426, 2008.

\bibitem{Szostak03}
J.~W. Szostak.
\newblock Functional information: molecular messages.
\newblock {\em Nature}, 423:689, 2003.

\bibitem{Quastler53}
H.~Quastler (Ed.).
\newblock {\em Essays on the use of information theory in biology}.
\newblock University of Illinois, Urbana, Illinois, 1953.

\bibitem{Rashevsky55}
N.~Rashevsky.
\newblock Life, information theory, and topology.
\newblock {\em Bull. Math. Bio.}, 17:229--235, 1955.

\bibitem{Atlan72}
H.~Atlan.
\newblock {\em L'organisation biologique et la th\'eorie de l'information}.
\newblock Hermann, 1972.

\bibitem{Berger03}
T.~Berger.
\newblock Living information theory.
\newblock {\em IEEE Info. Theory Soc. Newsletter}, 53:1--19, 2003.

\bibitem{Adami04}
C.~Adami.
\newblock Information theory in molecular biology.
\newblock {\em Physics of Life Reviews}, 1:3--22, 2004.

\bibitem{Taylor07}
S.~F. Taylor, N.~Tishby, and W.~Bialek.
\newblock Information and fitness.
\newblock arXiv:0712.4382, 2007.

\bibitem{Polani09}
D.~Polani.
\newblock Information: currency of life?
\newblock {\em HFSP Journal}, 5:307--316, 2009.

\bibitem{Shannon48}
C.~E. Shannon.
\newblock A mathematical theory of communication.
\newblock {\em Bell System Tech. Journal}, 27:379--423, 1948.

\bibitem{Shannon59}
C.~E. Shannon.
\newblock Coding theorems for a discrete source with a fidelity criterion.
\newblock {\em IRE National Convention Record}, 7:142--163, 1959.

\bibitem{CoverThomas91}
T.~M. Cover and J.~A. Thomas.
\newblock {\em Elements of information theory}.
\newblock Wiley-Interscience, New-York, 1991.

\bibitem{Csiszar08}
I.~Csisz\'ar.
\newblock Axiomatic characterizations of information measures.
\newblock {\em Entropy}, 10:261--273, 2008.

\bibitem{Shannon56}
C.~Shannon.
\newblock The bandwagon.
\newblock {\em Trans. Info. Theory}, 2:3--3, 1956.

\bibitem{RosenbluethWienerBigelow43}
A.~Rosenblueth, N.~Wiener, and J.~Bigelow.
\newblock Behavior, purpose and teleology.
\newblock {\em Philosophy of Science}, 10(1):18--24, 1943.

\bibitem{Wiener48}
N.~Wiener.
\newblock {\em Cybernetics: or Control and Communication in the Animal and the
  Machine}.
\newblock The MIT Press, 1948.

\bibitem{Ashby58}
W.~R. Ashby.
\newblock Requisite variety and its implications for the control of complex
  systems.
\newblock {\em Cybernetica}, 1:83--99, 1958.

\bibitem{Ashby56}
W.~R. Ashby.
\newblock {\em An introduction to cybernetics}.
\newblock Chapman \& Hall Ltd, 1956.

\bibitem{TouchetteLLoyd00}
H~Touchette and S~Lloyd.
\newblock Information-theoretic limits of control.
\newblock {\em Phys Rev Lett}, 84(6):1156--1159, 2000.

\bibitem{TouchetteLloyd04}
H.~Touchette and S.~Lloyd.
\newblock Information-theoretic approach to the study of control systems.
\newblock {\em Physica A}, 331:140--172, 2004.

\bibitem{Mitter01}
S.~K. Mitter.
\newblock Control with limited information.
\newblock {\em European J. Control}, 7(2-3):122--131, 2001.

\bibitem{Massey90}
J.~L. Massey.
\newblock Causality, feedback and directed information.
\newblock {\em Proc. Intl. Symp. Info. Theory Applic. (ISITA-90)}, pages
  303--305, 1990.

\bibitem{Marko73}
H.~Marko.
\newblock The bidirectional communication theory - a generalization of
  information theory.
\newblock {\em IEEE Trans. Info. Theory}, 21:1345--1351, 1973.

\bibitem{SegerBrockmann87}
J.~Seger and H.~J. Brockmann.
\newblock What is bet-hedging?
\newblock {\em Oxford Surveys in Evolutionary Biology}, 4:182--211, 1987.

\bibitem{LewontinCohen69}
R.~C. Lewontin and D.~Cohen.
\newblock On population growth in a randomly varying environment.
\newblock {\em Proc. Nat. Acad. Sci. USA}, 62:1056--1060, 1969.

\bibitem{Real80}
L.~A. Real.
\newblock Fitness uncertainty and the role of diversification in evolution and
  behaviour.
\newblock {\em Am. Nat.}, 115:623--638, 1980.

\bibitem{Stearns00}
S.~C. Stearns.
\newblock Daniel {B}ernoulli (1738): evolution and economics under risk.
\newblock {\em J. Biosci.}, 25:221--228, 2000.

\bibitem{Wagner03}
A.~Wagner.
\newblock Risk management in biological evolution.
\newblock {\em J. Theor. Biol.}, 225:45--57, 2003.

\bibitem{Stephens89}
D.~W. Stephens.
\newblock Variance and the value of information.
\newblock {\em Am. Nat.}, 134:128--140, 1989.

\bibitem{BergstromLachmann04}
C.~T. Bergstrom and M.~Lachmann.
\newblock Shannon information and biological fitness.
\newblock In {\em Information theory workshop IEEE '04 (San Antonio, Texas)},
  pages 50--54, 2004.

\bibitem{KussellLeibler05}
E.~Kussell and S.~Leibler.
\newblock Phenotypic diversity, population growth, and information in
  fluctuating environments.
\newblock {\em Science}, 309:2075--2078, 2005.

\bibitem{Donaldson10}
M.~C. Donaldson-Matasci, C.~T. Bergstrom, and M.~Lachmann.
\newblock The fitness value of information.
\newblock {\em Oikos}, 119:219--230, 2010.

\bibitem{Kelly56}
J.~Kelly.
\newblock New interpretation of information rate.
\newblock {\em Bell Syst. Tech. J.}, 35:917--926, 1956.

\bibitem{AlgoetCover88}
P.~H. Algoet and T.~M. Cover.
\newblock Asymptotic optimality and asymptotic equipartition properties of
  log-optimum investment.
\newblock {\em Ann. Prob.}, 16:876--898, 1988.

\bibitem{BarronCover88}
A.~R. Barron and T.~M. Cover.
\newblock A bound on the financial value of information.
\newblock {\em IEEE Trans. Inform. Theory}, 34:1097--1100, 1988.

\bibitem{Cover98}
T.~M. Cover.
\newblock Shannon and investment.
\newblock {\em IEEE Info. Theory Soc. Newsletter}, (Special Golden Jubilee
  Issue):10--11, 1998.

\bibitem{Breiman61}
L.~Breiman.
\newblock Optimal gambling systems for favorable games.
\newblock In {\em Fourth Berkeley symposium on mathematical statistics and
  probability}, pages 65--78, University of California Press, Berkeley, 1961.

\bibitem{PermuterKim08}
H.~H. Permuter, Y.-H. Kim, and T.~Weissman.
\newblock On directed information and gambling.
\newblock In {\em Proc. International Symposium on Information Theory (ISIT),
  Toronto, Canada}, 2008.

\bibitem{PerkinsSwain09}
T.~J. Perkins and P.~S. Swain.
\newblock Strategies for cellular decision-making.
\newblock {\em Mol. Syst. Bio.}, 5:326, 2009.

\bibitem{KarlinTaylor75}
S.~Karlin and H.~M. Taylor.
\newblock {\em A first course in stochastic processes}.
\newblock Academic Press, 1975.

\bibitem{Bernoulli38}
D.~Bernoulli.
\newblock Specimen theoriae novae de mensura sortis.
\newblock {\em Papers Imp. Acad. Sci. St. Petersburg}, 5:175--192, 1738.

\bibitem{Markowitz52}
H.~Markowitz.
\newblock Portfolio selection.
\newblock {\em The Journal of finance}, 7:77--91, 1952.

\bibitem{Mosegaard05}
M.~Morsegaard Christensen.
\newblock On the history of the growth optimal portfolio.
\newblock Unpublished, 2005.

\bibitem{Samuelson71}
P.~A. Samuelson.
\newblock The fallacy of maximaizing the geometric mean in long sequences of
  investing or gambling.
\newblock {\em Proc. Nat. Acad. Sci. USA}, 68:2493--2496, 1971.

\bibitem{MaynardSmithPrice73}
J.~Maynard Smith and G.~Price.
\newblock The logic of animal conflict.
\newblock {\em Nature}, 248:15--18, 1973.

\bibitem{MillsBeatty79}
S.~Mills and J.~Beatty.
\newblock The propensity interpretation of fitness.
\newblock {\em Philosophy of Science}, 46:263--286, 1979.

\bibitem{BeattyFinsen89}
J.~Beatty and Finsen S.
\newblock Rethinking the propensity interpretation: a peek inside {P}andora's
  box.
\newblock In M.~Ruse, editor, {\em What the philosophy of biology is, Essays
  dedicated to {D}avid {H}ull}. Kluwer Academic Publishers, 1989.

\bibitem{Robson96}
A.~J. Robson.
\newblock A biological basis for expected and non-expected utility.
\newblock {\em J. Econ. Theory}, 68:397--424, 1996.

\bibitem{FurstenbergKesten60}
H.~Furstenberg and H.~Kesten.
\newblock Products of random matrices.
\newblock {\em Ann. Math. Statist.}, 31:457--469, 1960.

\bibitem{Kingman73}
J.~F.~C. Kingman.
\newblock Subadditive ergodic theory.
\newblock {\em Ann. Prob.}, 1:883--899, 1973.

\bibitem{Tanny81}
D.~Tanny.
\newblock On multitype branching processes in a random environment.
\newblock {\em Adv. Appl. Prob.}, 13:464--497, 1981.

\bibitem{Hartley28}
R.~V.~L. Hartley.
\newblock Transmission of information.
\newblock {\em Bell System Tech. Journal}, page 535, 1928.

\bibitem{Kramer98}
G.~Kramer.
\newblock {\em Directed information for channels with feedback}.
\newblock PhD thesis, Swiss Federal Institute of Technology (ETH), Zurich,
  1998.

\bibitem{Kim08}
Y.-H. Kim.
\newblock A coding theorem for a class of stationary channels with feedback.
\newblock {\em IEEE Trans. Info. Theory}, 25:1488--1499, 2008.

\bibitem{Witsenhausen71}
H.~S. Witsenhausen.
\newblock Separation of estimation and control for discrete time systems.
\newblock {\em Proc. IEEE}, 59:1557--1566, 1971.

\bibitem{Gastpar03}
M.~Gastpar, B.~Rimoldi, and M.~Vetterli.
\newblock To code, or not to code: lossy source-channel communication
  revisited.
\newblock {\em IEEE Trans. Info. Theory}, 49:1147--1158, 2003.

\bibitem{KussellLeibler06}
E.~Kussell, S.~Leibler, and A.~Grosberg.
\newblock Polymer-population mapping and localization in the space of
  phenotypes.
\newblock {\em Phys. Rev. Lett.}, 97:068101, 2006.

\bibitem{IyengarCover00}
G.~N. Iyengar and T.~M. Cover.
\newblock Growth optimal investment in horse race markets with costs.
\newblock {\em IEEE Trans. Info. Theory}, 46:2675--2683, 2000.

\bibitem{Tulja03}
S.~Tuljapurkar, C.~C. Horvitz, and J.~B. Pascarella.
\newblock The many growth rates and elasticities of populations in random
  environments.
\newblock {\em Am. Nat.}, 162:489--502, 2003.

\bibitem{HaccouIwasa95}
P.~Haccou and Y.~Iwasa.
\newblock Optimal mixed strategies in stochastic environments.
\newblock {\em Theor. Pop. Biol.}, 47:212--243, 1995.

\bibitem{SasakiEllner95}
A.~Sasaki and S.~Ellner.
\newblock The evolutionarily stable phenotype distribution in a random
  environment.
\newblock {\em Evolution}, 49:337--350, 1995.

\bibitem{Merton69}
R.~C. Merton.
\newblock Lifetime portfolio selection under uncertainty: the continuous-time
  case.
\newblock {\em Rev. Eco. Stat.}, 51:247--257, 1969.

\bibitem{GodfreySmith07}
P.~Godfrey-Smith.
\newblock Information in biology.
\newblock In D.~Hull and M.~Ruse (eds.), editors, {\em The Cambridge Companion
  to the Philosophy of Biology}, pages 103--119. Cambridge University Press,
  2007.

\bibitem{Pack98}
L.~Pack Kaelbling, M.~L. Littman, and A.~R. Cassandra.
\newblock Planning and acting in partially observable stochastic domains.
\newblock {\em Artificial Intelligence}, 101:99--134, 1998.

\bibitem{Berger85}
J.~O. Berger.
\newblock {\em Statistical decision theory and {B}ayesian analysis}.
\newblock Springer-Verlag, 1985.

\bibitem{Leslie45}
P.~Leslie.
\newblock On the use of matrices in certain population mathematics.
\newblock {\em Biometrika}, 33:183--212, 1945.

\bibitem{AthreyaNey72}
K.~B. Athreya and P.~E. Ney.
\newblock {\em Branching processes}.
\newblock Springer-Verlag, 1972.

\bibitem{KurtzLyons97}
T.~G. Kurtz, R.~Lyons, R.~Pemantle, and Y.~Peres.
\newblock A conceptual proof of the {K}esten-{S}tigum theorem for multi-type
  branching processes.
\newblock In K.~B. Athreya and P.~Jagers, editors, {\em Classical and Modern
  Branching Processes}, volume~84 of {\em IMA Volumes in Mathematics and its
  Applications}, pages 181--185. Springer-Verlag, New York, 1997.

\end{thebibliography}
\end{document}